\documentclass[aps,reprint,amsmath,amssymb,longbibliography, superscriptaddress]{revtex4-2}

\PassOptionsToPackage{bbgreekl}{mathbbol}

\usepackage{ajtex-revtex}

\makeatletter 
    
\renewcommand\onecolumngrid{
\do@columngrid{one}{\@ne}%
\def\set@footnotewidth{\onecolumngrid}
\def\footnoterule{\kern-6pt\hrule width 1.5in\kern6pt}%
}

\renewcommand\twocolumngrid{
\def\footnoterule{
\dimen@\skip\footins\divide\dimen@\thr@@
\kern-\dimen@\hrule width.5in\kern\dimen@}
\do@columngrid{mlt}{\tw@}
}%

\makeatother

\renewcommand{\paragraph}[1]{\vspace{1em}\noindent\textbf{#1}}

\definecolor{ogreen}{RGB}{71,191,145}
\newcommand{\rl}[1]{\textcolor{black}{#1}}
 
\newcommand{\aj}[1]{\textcolor{black}{#1}} 
 
\newcommand{\op}[1]{\textcolor{black}{#1}}

\def\Ampere{Amp\`ere}

\begin{document} 

\title{Resistive relativistic magnetohydrodynamics without Amp\`ere's Law}

\author{Ruben Lier}\email{rlier@uva.nl}
\author{Akash Jain}\email{ajain@uva.nl}

\affiliation{Institute for Theoretical Physics, University of Amsterdam, 1090 GL Amsterdam, The Netherlands}
\affiliation{Dutch Institute for Emergent Phenomena, 1090 GL Amsterdam, The Netherlands}
\affiliation{Institute for Advanced Study, University of Amsterdam, Oude Turfmarkt 147, 1012 GC Amsterdam, The Netherlands}
\author{Jay Armas}\email{j.armas@uva.nl}

\affiliation{Institute for Theoretical Physics, University of Amsterdam, 1090 GL Amsterdam, The Netherlands}
\affiliation{Dutch Institute for Emergent Phenomena, 1090 GL Amsterdam, The Netherlands}
\affiliation{Institute for Advanced Study, University of Amsterdam, Oude Turfmarkt 147, 1012 GC Amsterdam, The Netherlands}
\affiliation{Niels Bohr International Academy, The Niels Bohr Institute, University of Copenhagen,
Blegdamsvej 17, DK-2100 Copenhagen \O{}, Denmark}

\author{Oliver Porth}\email{o.j.g.porth@uva.nl}
\affiliation{Anton Pannekoek Institute, Science Park 904, 1098 XH, Amsterdam, The Netherlands}

\date{\today}

\begin{abstract}
Resistive magnetohydrodynamics is thought to play a key role in transient high-energy astrophysical phenomena such as flares from black hole and neutron star magnetospheres. When performing numerical simulations of resistive magnetohydrodynamics, one is faced with the issue that Ampère's law becomes stiff in the high conductivity limit which poses challenges to the numerical evolution. We show that using a description of resistive magnetohydrodynamics based on higher-form symmetry, one can perform simulations with a generalized dual Faraday tensor without having to use Ampère's Law, thereby avoiding the stiffness problem. We also explain the relation of this dual model to a traditional description of resistive magnetohydrodynamics and how causality is guaranteed by introducing second order corrections.
\end{abstract} 

\maketitle

\section{Introduction}
Relativistic magnetohydrodynamics (MHD) provides a powerful means to study and model diverse high-energy astrophysical phenomena \cite{PorthChatterjeeEtAl2019a,Moesta2020,LiskaKaazEtAl2023,CosimoBambiSwarnimShashankEtAl2025}.  While the broad-brush global dynamics is well described by dissipationless ``ideal'' MHD, the formation of small scale (turbulent) structures inevitably leads to localized dissipation sites.  A well-known example are reconnecting current-sheets, which are able to drive high-energy particle acceleration leading to non-thermal transient emission \cite{NalewajkoGiannios2011,SironiSpitkovsky2014,LyutikovSironi2017a,ComissoSironi2019,YangLiEtAl2020a,Ripperda_2022,MahlmannPhilippovEtAl2023}. Ultimately, in the small-scale current-sheets, magnetic reconnection is required to release magnetic energy and change the topology of the magnetic field lines. However, as dictated by Alfven's theorem \cite{Davidson_2001}, magnetic reconnection only occurs when one accounts for finite plasma resistivity. Thus, while the magnetic Reynolds number (which signifies the ratio between system- and resistive-scales) of many astrophysical systems is extremely large \cite{GoedbloedPoedts2004}, when zooming into current sheets to study microscopic effects, resistivity needs to be included in the model explicitly~\cite{Del-ZannaPapini2016, RipperdaBacchiniEtAl2019, SelviPorthEtAl2023, BugliLoprestiEtAl2024}. To do so, in the relativistic case, one traditionally introduces the electric field as a nonhydrodynamic variable, whose lifetime is set by the conductivity that enters as a damping term in Amp\`ere's law, thereby introducing stiffness to the equations. This stiffness forces the time-step in explicit simulations to be extremely small which can be avoided by considering more complex numerical schemes, namely the implicit-explicit solver method (IMEX) \cite{Palenzuela_2009}. 

In this work, we take a different approach to solving the issue of magnetic dissipation, which involves thinking differently about what resistive MHD means compared to its traditional formulation. Electromagnetism has a \emph{one-form global symmetry}~\cite{Gaiotto_2015} that is responsible for enforcing the conservation of magnetic field lines as the associated string-like Noether charge. Therefore, instead of viewing resistive MHD as a limit of relativistic hydrodynamics coupled to electromagnetism, we can interpret it as a dissipative fluid with one-form symmetry, termed \emph{one-form MHD}~\cite{schubring, Grozdanov_2017, Hernandez_2017,Armas:2018atq, Armas20201}. This dual viewpoint entirely circumvents the pathological characteristic of resistive MHD because, instead of generating reconnection through a large relaxation term, resistivity is introduced as a small coefficient which forms the magnetic equivalent of viscosity. We show this by developing a numerical scheme for simulating astrophysical plasmas based on a causal model that avoids stiffness and is able to handle strong shocks at velocities close to the speed of light. 

\section{Dissipative MHD as one-form fluid}%
When describing relativistic plasmas involved in astrophysical phenomena, the relevant conserved hydrodynamic quantities are four-momentum and mass density, with the respective stress-energy tensor $T^{\mu\nu}$ and mass current $\rho^\mu$. Furthermore, a magnetohydrodynamic plasma has a dynamical magnetic field, which can be viewed as a one-form ``string density''. The associated conserved current is the dual Faraday's tensor $J^{\mu\nu}=\half\epsilon^{\mu\nu\rho\sigma}F_{\rho\sigma}$, which is conserved due to the Bianchi identity, where $F_{\mu\nu}$ is the electromagnetic field strength tensor. The conservation laws are thus given by
\begin{align} \label{eq:conservationlaws}
    \nabla_{\nu} T^{\mu \nu }  =  0 ~~ , \quad
    \nabla_{\nu} J^{\mu \nu }   =  0 ~~ , \quad
    \nabla_{\nu} \rho^{\nu}   =  0  ~~ . 
\end{align}
Here $\nabla_\mu$ denotes the covariant derivative associated with the background spacetime metric $g_{\mu\nu}$.

It is common to assume a large conductivity of the plasma in its co-moving frame, which allows us to eliminate the electric field components as $\mathbf{E}=\mathbf{B}\times \mathbf{v}$, 
where $v^i$ is the fluid velocity in units of the speed of light, $E_i = F_{it}$ is the electric field, and $B^{i}=\epsilon^{ijk}F_{jk}$ is the magnetic field.  
The resulting system of conservation laws is known as ``ideal MHD'', which ensures that magnetic field lines are strictly topological (in the sense that they maintain their connectivity throughout the evolution and ``move with the flow'' \cite{GoedbloedPoedts2004}). The constitutive relations for ideal MHD are given as
\begin{subequations} \label{eq:idealconstitutiveequations1}
\begin{align}
    T_{(0)}^{\mu \nu } 
    & = \Big( \epsilon + p + b^2\Big) u^{\mu } u^{\nu} 
    +  \lb p + \half b^2\rb g^{\mu\nu}  
    - b^{\mu } b^{\nu }, \\   \label{eq:Jcurrent1}
    J_{(0)}^{\mu\nu} &= 2  u^{[\mu} b^{\nu] } ~~ ,  \\ 
    \rho_{(0)}^{\mu} &= \rho\, u^{\mu} ~~ , 
\end{align}
\end{subequations}
Here $u^\mu = \Gamma(1,{\bf v})$ is the fluid four-velocity, with the Lorentz factor $\Gamma=1/\sqrt{1-{\bf v}^2}$, and $b^\mu = \Gamma({\bf B}\cdot{\bf v},{\bf B}-{\bf v}\times{\bf E})$ is the magnetic field in the fluid rest frame with $b^2=b^\mu b_\mu$. Note that $b^\mu u_\mu=0$ and $u^\mu u_\mu=-1$. Furthermore, $\epsilon$ is the total fluid energy density (including the rest-mass density $\rho$) and $p$ is the fluid pressure. The thermodynamic relations are given as
\begin{align} \label{eq:thermodynamicrelations}
    \epsilon + p 
    =  Ts + \mu\rho~~, \qquad
    \df p =  s\, \df T  + \rho\, \df\mu  ~~ , 
\end{align}
where $T$ is temperature, $s$ is entropy density, and $\mu$ is the mass chemical potential. In writing \cref{eq:idealconstitutiveequations1}, we have assumed that magnetic fields decouple from the fluid equation of state, i.e. the total thermodynamic pressure decouples into $P(T,\mu,b^2) = p(T,\mu) + b^2/2$.
We close the equations by specifying the equation of state, which is taken to be that of an ideal gas given by
\begin{align}\label{eq:EoS}
    \epsilon = \rho + \frac{p}{\hat{\gamma} - 1}~~,
\end{align}
where $\hat{\gamma}$ is the adiabatic index. See~\cite{Del_Zanna_2007,GoedbloedPoedts2004} for more details on ideal MHD.

As mentioned in the introduction, finite conductivity effects are necessary in order for magnetic field lines to diffuse and reconnect. In the traditional formulation of resistive MHD, one accounts for this by invoking the \Ampere's law that imparts relaxational dynamics to electric fields \footnote{Normally, the conductivity term in \eqref{eq:ampere} should also contain charge diffusion, i.e. $e_\nu - T\partial_\nu(\mu_q/T)$ instead of $e_\nu$, where $\mu_q$ is the chemical potential conjugate to $q$. However, since $q=-u_\mu\nabla_\nu F^{\mu\nu}\sim\cO(\dow)$, the diffusion term is subleading in derivatives and has thus been ignored.}, i.e.
\begin{align}\label{eq:ampere}
    \nabla_\nu F^{\mu\nu} =  q\, u^{\mu} 
    + \Big(\sigma_{\parallel} \hat b^{\mu} \hat  b^{\nu} + \sigma_{\perp} \mathbb{B}^{\mu \nu  } \Big)  e_{\nu},
\end{align}
where $e^\mu = \Gamma ({\bf E}\cdot{\bf v},{\bf E}+{\bf v}\times {\bf B})$ is the electric field in the fluid rest frame, $q$ is the charge density, and $\sigma_\|$, $\sigma_\perp$ are the longitudinal and transverse conductivities. Furthermore, $\hat b^{\mu} = b^{\mu}/|b|$ and $ \mathbb{B}^{\mu\nu} = g^{\mu\nu} + u^{\mu } u^{\nu} - \hat b^{\mu} \hat b^{\nu}$. See more details in App.~\ref{eq:traditionalMHD}. Note that $J^{\mu\nu}=2u^{[\mu}b^{\nu]}+\epsilon^{\mu\nu\rho\sigma}u_\rho e_\sigma$. 

In this work, we instead include resistivity in a way that avoids using Ampère's law and the electric field but instead treats $J^{\mu \nu }$ as a general current with dissipative corrections that account for diffusion of magnetic field lines. \cite{Grozdanov_2017,schubring}. To wit, 
\begin{align} \label{eq:Jexpansion}
        J^{\mu \nu }   = J_{(0)}^{\mu \nu }   + J^{\mu \nu }_{(1)}  + \mathcal{O} (\partial^2 ) ~~ . 
\end{align}

\op{It is worth pointing out that since $J_{\mu \nu }$ is fundamentally defined as $\frac{1}{2}\epsilon_{\mu \nu \rho \sigma } F^{\rho \sigma }$, the scalar $F_{\mu \nu } J^{\mu \nu }$ a relativistic invariant also when $J_{\mu \nu }$ is expanded in this way \cite{Armas20201}.}

For simplicity, we will keep $T^{\mu\nu}$ and $\rho^\mu$ as ideal. We then consider the thermodynamic relations of \eqref{eq:thermodynamicrelations} to obtain the dissipation rate
\begin{align}  \label{eq:secondlaw}
    \nabla_{\mu} S^{\mu}   =   -  J^{\mu \nu}_{(1) } \left(  \partial_{\mu} \frac{b_{\nu} }{T} \right) + \mathcal{O}(\partial^3) = \Delta  ~~ . 
\end{align}
where $S^{\mu}$ is the canonical entropy current given by
\begin{align}
\begin{split}
       T  S^{\mu}   & = p u^{\mu} - b_{\nu}  J^{\mu \nu }   -  u_{\nu } T^{\mu \nu } - \mu \rho^{\mu}   \\ 
        & = T s u^{\mu}  -  b_{\nu}  J_{(1) }^{\mu \nu }    ~~ . 
        \end{split}
\end{align}
Working in a frame where the magnetic field is unaffected by dissipative corrections, there are only two first-order terms that can be constructed that are in accordance with the second law of thermodynamics $ \Delta  \geq 0 $, to wit
\begin{align}  \label{eq:generalcurrent}
    J^{\mu \nu}_{(1) }    &  =  -   \left(  2 r_{\perp }  \mathbb{B}^{ \rho  [ \mu } \hat b^{\nu ] } \hat  b^{\sigma}   
 + r_{\parallel }  \mathbb{B}^{\mu \rho }  \mathbb{B}^{\nu \sigma } \right) 2   T \partial_{[ \rho } \left(  \frac{b_{\sigma ] } }{T} \right), 
\end{align}
where $\hat b^{\mu}  = b^{\mu} / |b| $ and $ \mathbb{B}^{\mu \rho }  = \eta^{\mu \nu } + u^{\mu } u^{\nu} - \hat b^{\mu} \hat b^{\nu}$. The second law of thermodynamics requires
\begin{align}
    r_{\parallel} \geq  0 ~~ , ~~ r_{\perp} \geq 0  ~~ . 
\end{align}
\rl{The coefficients $r_{\parallel,\perp}$ are named after resistivity. This is because in the simple ultra-relativistic limit, up to a factor that involves the enthalpy density and the magnetic field squared, one can express $r_{\parallel,\perp}$ in \eqref{eq:generalcurrent} in terms of the inverse of the conductivity of traditional resistive MHD \cite{Armas20201,Hernandez_2017} (see App.~\ref{eq:transform}).}

\section{Stability and causality}

Including the dissipative terms in \eqref{eq:generalcurrent} means that the evolution of magnetic field lines becomes diffusive. Such diffusive modes violate causality which lead to instabilities when the fluid velocity approaches the speed of light \cite{HISCOCK1983466,hiscock,PhysRevX.12.041001}. One option to restore causality is to follow the M\"uller-Israel-Stewart (MIS) prescription~\cite{ISRAEL1979341,Muller:1967zza} and add new gapped tensor fields to regulate the instabilities along the lines of~\cite{Chandra:2015iza, Cordeiro:2023ljz}. We instead implement the recently-discovered Bemfica-Disconzi-Noronha-Kovtun (BDNK) prescription \cite{Kovtun_2019,bemfica2022firstorder,Armas_2022,hoult2024} by including a second-derivative correction to \eqref{eq:Jexpansion} which turns the magnetic diffusive equation into a Telegrapher's equation, rendering it causal. We take
\begin{align}\label{eq:diffusiveterm}
    J^{\mu \nu }  
    = J^{\mu \nu }_{(0)}  
    + J^{\mu \nu }_{(1) }  
    - 2 \tau  u^{[\mu}  \nabla_{\rho} J_{(0)}^{\nu]  \rho }   
    + \mathcal{O} (\partial^3 )  ~~ ,
\end{align}
where $\tau$ can be viewed as a relaxation time that turns the Bianchi identity into a wave equation at short wavelengths. Note that $\nabla_{\nu} J_{(0)}^{\mu \nu } =   - \nabla_{\nu} J_{(1)}^{\mu \nu }+\ldots = \cO(\dow^2)$. 

To verify that the $\tau$ correction in \cref{eq:diffusiveterm} makes the equations causal, we consider plane wave solutions $\propto \exp(- i \omega t + i  \mathbf{k}  \cdot \mathbf{x} )$ to the linearized equations and compute the front velocity
\begin{align}
W  (\theta ) =  \lim_{ \kappa \rightarrow \infty }  \frac{ \omega}{|{\bf k}|} ~~ ,
\end{align}
and imposing~\cite{Krotscheck1978}
\begin{align} \label{eq:causalityconstraints1234}
    \text{Re}\,W(\theta )   \leq 1 ~~ , \qquad 
    \text{Im}\,W(\theta )   =  0  ~~  , 
\end{align}
where $|\mathbf{k}|$ denotes the magnitude of the wavevector and $\theta=\cos^{-1}(\hat{\bf B}\cdot\hat{\bf k})$ is its angle with respect to the background magnetic field.

We assume for simplicity that $r\equiv r_{\perp} = r_{\parallel}$, so that $J^{\mu \nu }$ is given by
\begin{equation}  \label{eq:fullmodel1}
    J^{\mu \nu}  
    =  J^{\mu \nu }_{(0)}
    -  2  r   P^{\mu \rho } P^{\nu \sigma }   T \partial_{[ \rho } \!\left(\frac{b_{\sigma ] } }{T} \right) -   2 \tau  u^{[\mu}  \nabla_{\rho} J_{(0)}^{\nu]  \rho }   ~~  , 
\end{equation}
where $P^{\mu \rho }  = \eta^{\mu \nu } + u^{\mu } u^{\nu} $. Let us consider the Alfven channel, where the front velocities are given by
\begin{align}
     W_{\text{Alfven}}^2 (\theta ) = \frac{b_0^2  \cos^2 ( \theta )}{   b_0^2+w_0 } 
  + \frac{r}{\tau} ~~, 
\end{align}
which can be constrained to satisfy $ W_{\text{Alfven}}^2  \leq 1 $ for sufficiently large $\tau / r  $. To obtain the front velocity of the magnetosonic sector $W_{\text{magnetosonic}} (\theta)$, it is important that we relate temperature $T$ to the other hydrodynamic variables. This can be done using the ideal gas relation 
\begin{align}
    T \propto \frac{p}{\rho}~~ , 
\end{align}
where the proportionality coefficient drops out of \eqref{eq:fullmodel1}. In App.~\ref{app:causality}, we discuss the front velocity of $W_{\text{magnetosonic}}$ and its causal nature for large $\tau / r$. We also show that in the ultra-relativistic limit the model can lead to front velocities that exactly coincide with those found for traditional resistive MHD for a specific anisotropic choice of $r_\|,r_\perp$ in App.~\ref{sec:connecting_higher_form}. This demonstrates that the second-order correction in \eqref{eq:diffusiveterm} allows one to mimic the inherent hyperbolic nature of traditional resistive MHD, which is discussed in App.~\ref{eq:traditionalMHD}.

\section{Numerical scheme}%

We now outline a numerical scheme to solve \eqref{eq:conservationlaws} with the two-form current \eqref{eq:diffusiveterm} and the equation of state \eqref{eq:EoS}. For simplicity, we take $r \equiv r_{\perp} = r_{\parallel}$. Our implementation is added as a physics module to \texttt{BHAC} \citep{PorthOlivares2017,OlivaresPorthEtAl2019}, which offers various numerical schemes to solve conservation laws on arbitrary background metrics.  
First, to make the equations amenable to numerical integrations, we project onto time-like hypersurfaces. This enables us to decompose the currents into conserved variables $\mathbf{U}$, (geometric) sources $\mathbf{S}$ and fluxes $\mathbf{F}^i$. Specializing to the Minkowski metric, this leads to
\begin{subequations} \label{eq:fluxequation}
\begin{align}
    \partial_t \mathbf{U} + \partial_i \mathbf{F}^i = \mathbf{S} ~~ , 
\end{align}
with 
\begin{align}
  \mathbf{U} = \begin{pmatrix}
    \rho^t \\
    T^{tk}\\
    T^{tt} \\
    J^{tk} \\
    b^k
    \end{pmatrix}, \quad 
  \mathbf{F}^i = 
  \begin{pmatrix}
    \rho u^i \\
    T^{ik} \\
    T^{it} \\
    J^{ik} \\
    0
  \end{pmatrix}, \quad
  \mathbf{S} = 
  \begin{pmatrix}
    0 \\
    0 \\
    0 \\
    0 \\
    \dot{b}^k \\
  \end{pmatrix}~~,
\end{align}
\end{subequations}
where the $k$ index is understood to run over the columns. This is implemented in \texttt{BHAC} using a finite-volume discretization. Throughout this work, we adopt second-order total variation diminishing time-stepper and spatial reconstruction techniques  (in particular the ``Koren'' limiter). 

Note that, in addition to the conservation equations, we have extended \eqref{eq:fluxequation} with the trivial evolution for $b^i$ due to the source $\dot{b}^i$. To compute $\dot{b}^i$, we follow the approach of  \cite{PandyaMostEtAl2022}, which means that we decompose $J^{ti}$ into
\begin{align}  \label{eq:decomposedequation}
    J^{ti} 
    = J^{\prime ti}  
    + M^{i}_{~j} \frac{\partial b^j }{\partial t }
 ~~ ,  \end{align}
where $J^{\prime t i  }$ is obtained from $J^{t i  }$ after omitting all terms involving $\partial b^j/\partial t$. The matrix $M^{i}_{\, \, j} $ is given by
\begin{equation}\label{eq:mij}
    M^i_{~j} 
    = r \Big( (1-\Gamma^2) \delta^{i}_j + u^i u_j \Big)  
    - \tau \Big( u^iu_j - \Gamma^2 \delta^{i}_j \Big)~~.
\end{equation}
Note that it is due to $\tau$ that  $M^{i}_{\, \, j} $ is nonzero in the rest frame. Inverting \eqref{eq:decomposedequation} yields 
\begin{align}\label{eq:bdotsource}
  \dot{b}^i = (M^{-1})_{~j}^i \Big(J^{tj}- J^{\prime tj}\Big)~~. 
\end{align}
Furthermore, the $t$-component of the one-form conservation equation yields $\partial_i J^{it} = 0$, which is essentially the magnetic Gauss law $\mathbf{\nabla\cdot B} = 0$. 

Solving (\ref{eq:fluxequation}) requires knowledge of the state in the form of the so-called primitive variables 
\begin{align}
    \mathbf{P} = \Big(
    ~\rho~,~ u^i~,~ p~,~ J^{ti}~,~ b^i~\Big)~~,
\end{align}
\op{and we further define a set of auxiliary variables $\mathbf{A}$ such that
\begin{align}
    \mathbf{A} = \Big(
    ~ \dot{u}^i~,~ \dot{\epsilon}~,~ \dot{\rho}~
    \Big)~.
\end{align}
With these definitions, the fluxes are expressed as $\mathbf{F}^i=\mathbf{F}^i(\mathbf{P}(\mathbf{U}),\mathbf{A})$. }
Our strategy to solve the non-linear primitive variable recovery $\mathbf{P}(\mathbf{U})$ by means of standard Newton-Raphson and Newton-Krylov methods is described in App.~\ref{eq:primitivevariablerecovery}. \op{The set $\mathbf{A}$ is required since we focus on dissipative corrections to the magnetic field alone, meaning that there are no equivalent expressions to (\ref{eq:bdotsource}) for $\dot{u}^i$, $\dot{\rho}$ and $\dot{\epsilon}$. In this work, we approximate them instead by finite differencing the current and previous state in the time evolution.    }

During flux computation, gradients in the dissipative fluxes (\ref{eq:fullmodel1}) are evaluated by second-order central differencing the reconstructed interface states, yielding left-biased and right-biased fluxes. From these, we compute upwinded numerical fluxes using an approximate Riemann solver, see e.g. Section 2.9 of \cite{PorthOlivares2017} for details.
Similarly, we use unlimited second-order central differencing to obtain gradients for the source term (\ref{eq:bdotsource}).  
For simplicity, the characteristic velocities are set to the speed of light, mirroring the implementation of resistive general relativistic magnetohydrodynamics in \texttt{BHAC} described in \cite{RipperdaBacchiniEtAl2019}.  
\op{Thus our numerical timestep becomes
\begin{align}
    \Delta t = {\rm CFL}\,\times\, {\rm min}_i(\Delta x^i/c)~~,
\end{align}
which introduces the $\rm{CFL}$ factor chosen $<1$ for numerical stability.}
To ensure that $\partial_i J^{it}=0$ is maintained to machine precision, we have implemented the cell-centered flux-constrained transport algorithm (FCT) described by \cite{Toth2000} as well as the staggered constrained transport scheme due to \cite{BalsaraSpicer1999,PorthOlivares2017}.
\op{Regarding the conservation of electric charge $\nabla\cdot\mathbf{E}$, similar constraint transport algorithms have been presented \cite[e.g.][]{MignoneMattiaEtAl2019}.  However, numerical tests in resistive relativistic MHD have not demonstrated superior behavior of such algorithms \citep[see e.g. the discussion and references in][]{MignoneBertaEtAl2024}.  Therefore, and consistent with  common practise in ideal MHD algorithms, we do not enforce charge conservation at the discrete level.}

\section{Results}
\label{sec:results}

\subsection{Telegrapher's equation}

\begin{figure}[t]
    \centering
    \includegraphics[trim={0 0 0 0},clip,width=0.48\textwidth]{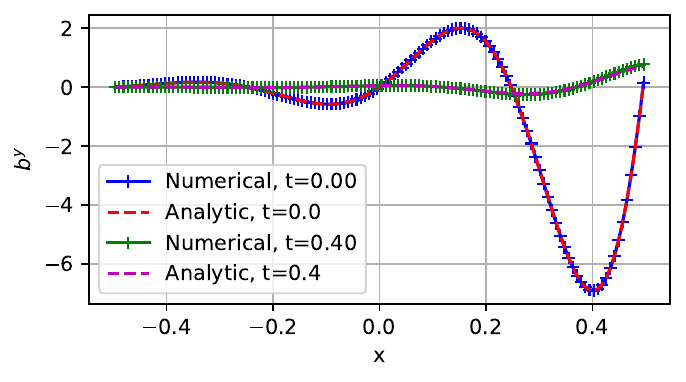}
    \caption{Boosted 1D telegraph solution comparing numerical and analytic solutions.  The numerical solution is obtained for 128 gridpoints and a CFL-number of 0.3.  }
    \label{fig:T1D}
\end{figure}

To verify that our numerical scheme is valid, we first perform an analytical benchmark in a simple limit. In the inert limit (e.g. $\rho,p\gg b^2$), the higher form MHD system reduces to the evolution of the magnetic field alone. Using the fact that the fluid velocity is non-dynamical \cite{10.1093/mnras/stae1729}, we go to the rest frame where $u^{\mu} = (1,0,0,0)$ and obtain
\begin{align}  \label{eq:Jcurrent}
    \partial_{i } J^{t  i }   &  =    \partial_{i}   b^{i }    + \tau   \partial_{i}         \partial_{t}  b^{ i}     =0   \implies  \partial_i b^{ i}    =0 ~, \\ 
    \begin{split}
              \partial_{t} J^{i  t  } +   \partial_{j} J^{i  j  }  &  =    - \partial_{t }   b^{ i }     - \tau   \partial^2_{t}      b^{  i  }          +     r            \partial^{ 2   } b^{ j  }         =0  ~~ , 
            \end{split} \label{eq:telegraphersequation}
 \end{align} 
 where \eqref{eq:telegraphersequation} is Telegrapher's equation. A similar Telegrapher's equation is recovered and used for benchmarking in the standard relativistic resistive MHD case \citep{MignoneMattiaEtAl2018,MignoneBertaEtAl2024}.  
To demonstrate the validity of the higher form MHD implementation, we first investigate the one-dimensional boosted solution given by 
\begin{align}
    b^{y}  (t, x )    &  =  \exp \left(-\frac{ 
    \Gamma  t  - u^x x }{2 \tau }\right)   \sin \left(\phi(t, x)\right)
\end{align}
where $\phi(t, x) = k (\Gamma  x  - u^x t ) -\Theta (\Gamma  t - u^x x) $ and we set $k=2\pi/L_x$, $L_x=1$, $r=\tau=0.1$, $u^x=1$, $\rho=10^{12}$, $p=10^{10}$.  
Given the analytic solution for $b^\mu(t,x)$, the initial condition for $J^{ti}$ can be obtained from (\ref{eq:diffusiveterm}).  
Since the boosted solution is non-periodic, the boundary conditions need to be obtained from the analytic solution and are continuously updated in time.  In Figure \ref{fig:T1D}, we show an exemplary solution at $t=0$ and $t=0.4$. 
As can be seen, the numerical realization shows excellent agreement with the analytic solution. 

\begin{figure}[t]
    \centering
    \includegraphics[trim={0 0 0 0},clip,width=0.48\textwidth]{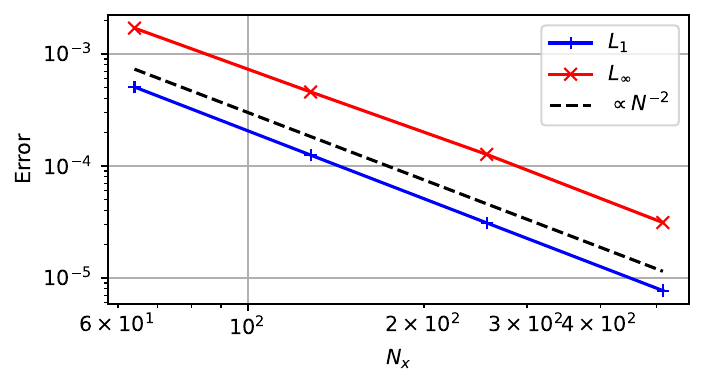}
    \caption{Convergence for the boosted and rotated 2D telegraph solution after simulating for $t=0.8$. }
    \label{fig:T2Dconvergence}
\end{figure}

To make this statement more quantitative and test a larger number of terms contributing to the time-evolution, we next compute the solution for a rotated 2D case which is boosted in $y$-direction.  
Again we set $\rho=10^{12}, p=10^{10}$ and the other non-trivial components read
\begin{align} \label{eq:solutionboosty12}
 b^{\prime  t }  (t, x^{\prime } , y^{\prime } )    &  =  u^y  \exp \left(-\frac{\Gamma   t  - u^y y^{\prime  } }{2 \tau}\right)   \sin \left(\phi(t, x', y')\right)~~, \nn\\ 
b^{\prime  y}  (t, x^{\prime } , y^{\prime }  )    &  =  \Gamma   \exp \left(-\frac{\Gamma  t - u^y y^{\prime  } }{2 \tau}\right)   \sin \left(\phi(t,x',y')\right)~~, \nn\\
u^{\prime y} & = u^y~~,
\end{align}
where $\phi(t, x', y') = k x^{\prime }  - \Theta  (\Gamma t  -  u^y y^{\prime })$.  Equation (\ref{eq:solutionboosty12}) is further rotated in the $xy$-plane by the angle $\alpha$. Hence, we apply the rotation matrix 
\begin{align}
    R =     
    \begin{pmatrix}
        1 & 0 & 0 & 0 \\ 
        0  & \cos(\alpha )  &  - \sin(\alpha )  &  0  \\ 
        0 & \sin(\alpha )  &  \cos(\alpha )   &  0 \\ 
        0 & 0 &  0  &  1 
    \end{pmatrix}~~,
\end{align}
to the $(\cdot)^{\prime}$ vector quantities and obtain the spatial coordinates from 
\begin{align}
\begin{split}
        x^{\prime}  &  =  \cos(\alpha ) x  +   \sin(\alpha )  y~~,  \\     
        y^{\prime}   & =  \cos(\alpha )  y   -  \sin(\alpha )  x  ~~ .
\end{split}
\end{align}
The numerical solution is obtained for $\tan\alpha=2$, $u^y=0.5$, $r=\tau=0.2$ in a computational domain $x\in[-1/2,1/2]$, $y\in[-1/4,1/4]$ for a range of grid points $N_x=2 N_y = [64,128,256,512]$.  We choose a timestep corresponding to a courant parameter of ${\rm CFL}=0.3$ and $\partial_i J^{it}=0$ preserving FCT algorithm.  The convergence against the analytic solution is demonstrated in Figure \ref{fig:T2Dconvergence} by means of the $L_1$ (mean error) and $L_\infty$ (maximum point-wise error) norms. As expected, the numerical scheme exhibits second order convergence in both norms, confirming the correctness of the implementation. 

\subsection{Shocktube tests}

\begin{figure*}[t]
    \centering
    \includegraphics[trim={0 0 0 0},clip,width=0.95\textwidth]{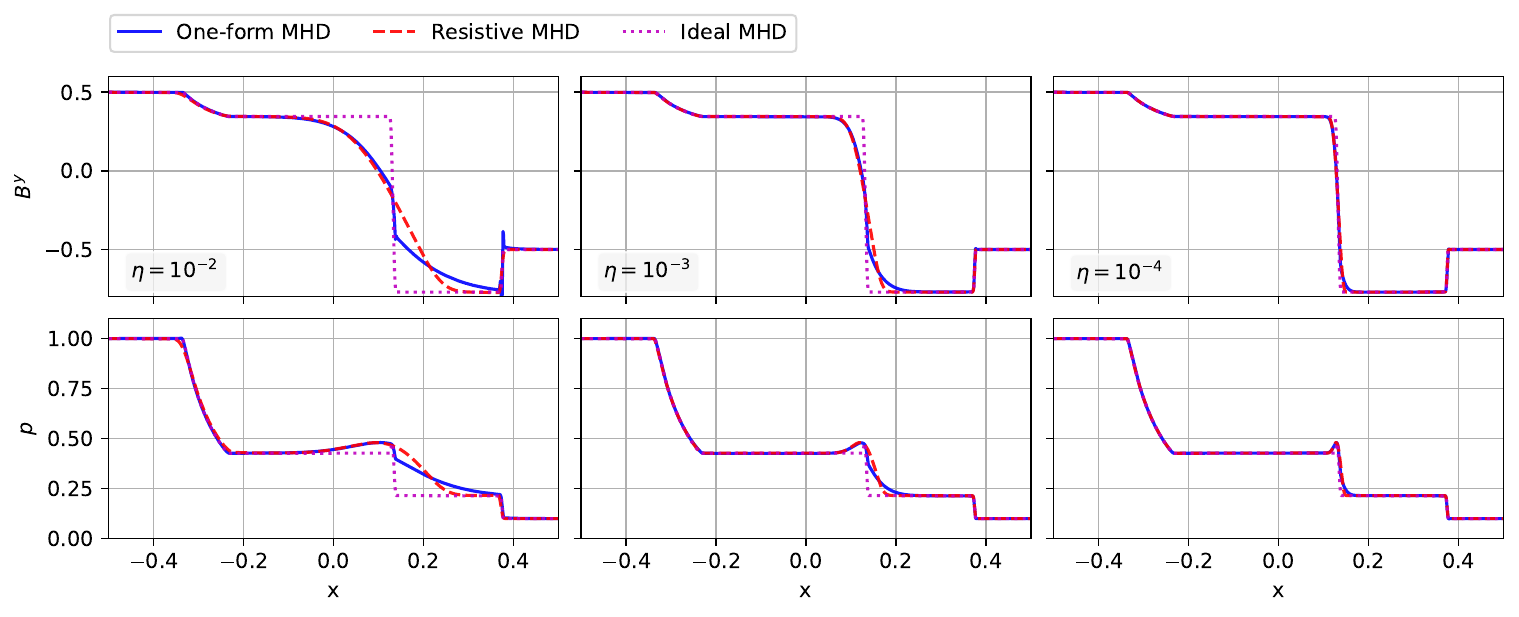}
    \caption{Shocktube problem for a range of resistivities $\eta$ compared with the ideal MHD case.  For the one-form MHD case we set $r=\eta$.  Shown are the out-of-plane magnetic field component (corresponding to $J^{ty}$ for one-form MHD) and the gas pressure in the top and bottom panel respectively.  Differences between the standard resistive MHD and one-form MHD are most noticeable near the tangential discontinuity for the high resistivity case $\eta=10^{-2}$.  Both schemes converge to the ideal solution which is closely realized at $\eta=10^{-4}$.}
    \label{fig:shocktube}
\end{figure*}

\op{Shock tubes are widely used one-dimensional tests to probe how numerical schemes fare in the presence of discontinuities \citep[e.g.][]{BrioWu1988}.  The initial condition simply consists of two states (left and right) which are separated by a discontinuity at $x=0$.  To compare with previous studies \cite{Palenzuela_2009, DionysopoulouAlic2013, Wright_2019}, we adopt here the following initial states:
\begin{align}
    \begin{split}
        \left(\rho^L,p^L,B^{y,L}\right) &= (1,1,1/2)~, \\
        \left(\rho^R,p^R,B^{y,R}\right) &= (1/8,1/10,-1/2)\, ,
    \end{split}
\end{align}
while all other quantities are initialized with zero and we set the adiabatic index to $\hat{\gamma}=2$.  This setup is adopted for ideal MHD, resistive MHD, as well as one-form MHD, where in the latter case we set $J^{ty}=B^y$.  We scan through a range of resistivities $\eta\in\{10^{-4},10^{-3},10^{-2}\}$ and identify $r=\eta$.  In general, we set the dampening timescale to $\tau=2\eta$, except for the near-ideal case where we needed to set $\tau=4\eta$ for numerical stability.  
The simulations are carried out with a resolution of 1024 cells in a domain of $x\in[-0.5,0.5]$ with a courant parameter of $\rm{CFL}=0.3$.  The results at the final time $t=0.4$ are shown in Figure \ref{fig:shocktube}.  }

\op{In this configuration, the solution consists of a left-going fast wave, a rarefaction, a perpendicular shock (at $x\simeq1.5$) and a right-going fast wave.  All models are able to handle the discontinuous initial data well and show no spurious oscillations.  
In this test, one-form MHD and fully resisistive MHD agree well for resistivities of $\eta=10^{-3}$ and below and both models closely resemble the ideal case at $\eta=10^{-4}$.  Only for the largest value of $\eta=10^{-2}$, there are systematic differences in the region between the shock and the right-going fast wave.  
It should be noted that such differences in the large $\eta$ limit are indeed expected as the resistive case approaches the electrovacuum limit for $\eta\to \infty$, whereas the electromagnetic fields remain coupled to the fluid in the one-form MHD case.  
However, for most applications such as magnetic reconnection and turbulence, the high-Lundquist number (thus low $\eta$) regime is more relevant and here one-form MHD shows very good correspondence to fully resistive MHD.  }

\subsection{Cylindrical explosion}

\begin{figure*}[t]
    \centering
    \includegraphics[trim={0 0 0 0},clip,width=0.85\textwidth]{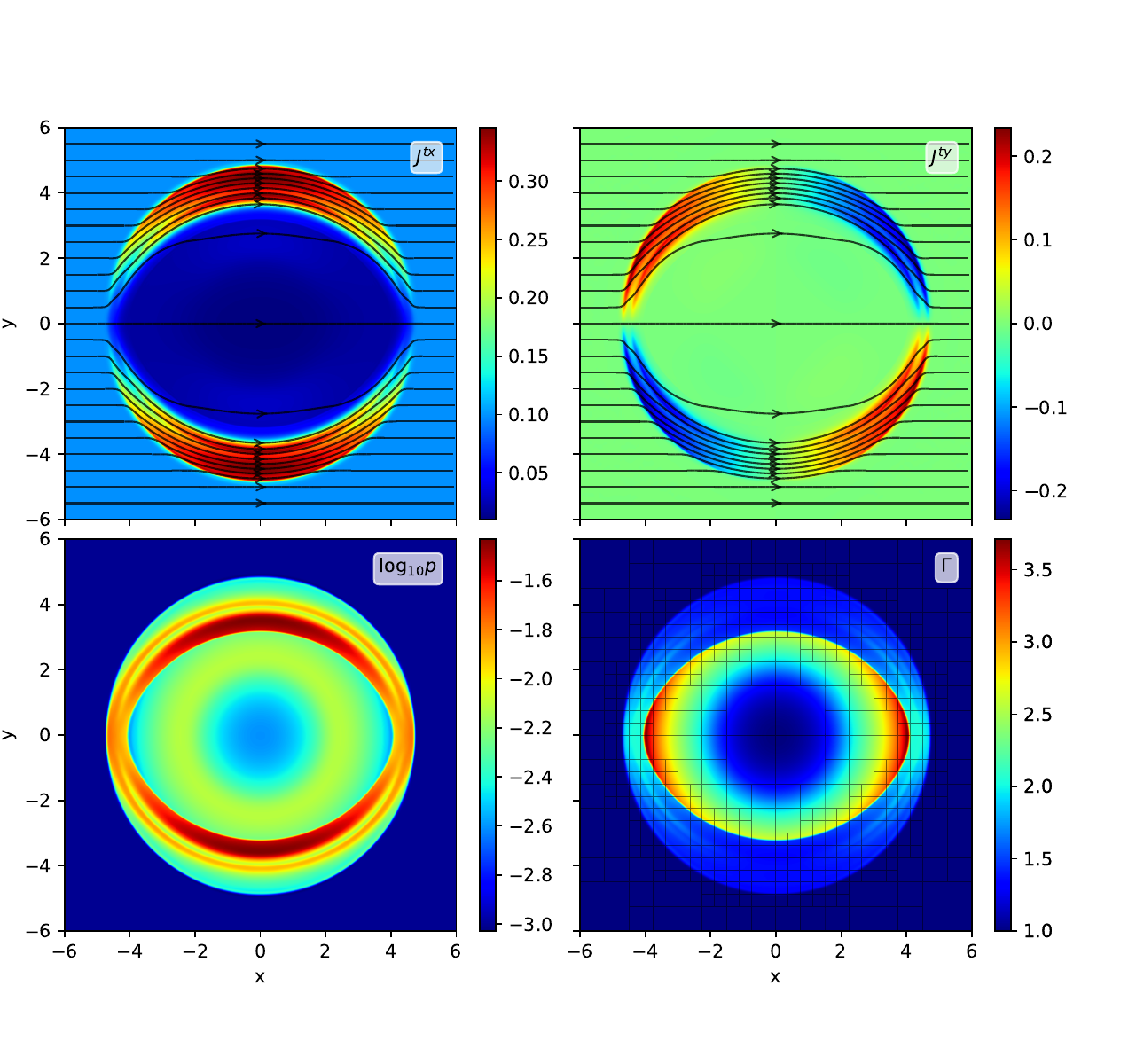}
    \caption{Strong cylindrical explosion test according to \citet{Komissarov2007} with $r=0.018,\, \tau=0.09$.  The top panels show the magnetic field components ($J^{tx},J^{ty}$) along with their fieldlines.  The bottom panels show $\log_{10}p$ (left) and the Lorentz factor $\Gamma$ (right).  In the latter panel we indicate the adaptive mesh blocks which consist of $16^2$ cells each. }
    \label{fig:explosion}
\end{figure*}

\op{
As a first multi-dimensional shock-dominated test, we carry out the cylindrical explosion with parameters identical to \citet{Komissarov2007}.  
The initial condition consists of a uniform magnetic field at rest $b^x=J^{tx}=0.1$, a background plasma with pressure and density of $p=\rho=0.001$ and an inner cylinder $r<0.8$ of high pressure $p=1$ and density $\rho=0.01$.  Just as in \cite{Komissarov2007}, we taper off $p$ and $\rho$ in the region $0.8<r<1$ by multiplication with $\exp(-(r-0.8))$.  
The domain extends over $x,y\in[-6,6]$ which we discretize by $128^2$ cells and allow for two additional levels of mesh refinement to reach an effective resolution of $512^2$ cells.  We adopt $r=0.018$, $\tau=5 r$ and set the adiabatic index to $\hat{\gamma}=4/3$.  Furthermore, the staggered mesh constrained transport scheme is used and we set the CFL parameter to $0.2$.  
}

\op{
The solution at time $t=4$ is shown in Figure \ref{fig:explosion}.  As expected, for this low $r$, the solution matches well with the ideal MHD solution \cite{Komissarov1999a} and the reference resistive case from \cite{Komissarov2007}.  Though similar to the shock tube case, upon closer inspection we note slight differences in the shocked region which are best visible by the outer orange ring in the logarithmic pressure plot.  A corresponding feature is seen in the magnetic field which shows a slight depression in this area.  In the purely resistive and ideal cases on the other hand, the shocked region remains monotonous.  
}

\subsection{Orszag-Tang vortex}

\begin{figure*}[t] 
    \centering
    \includegraphics[trim={0 1.1cm 1cm 0.5cm},clip,width=17cm]{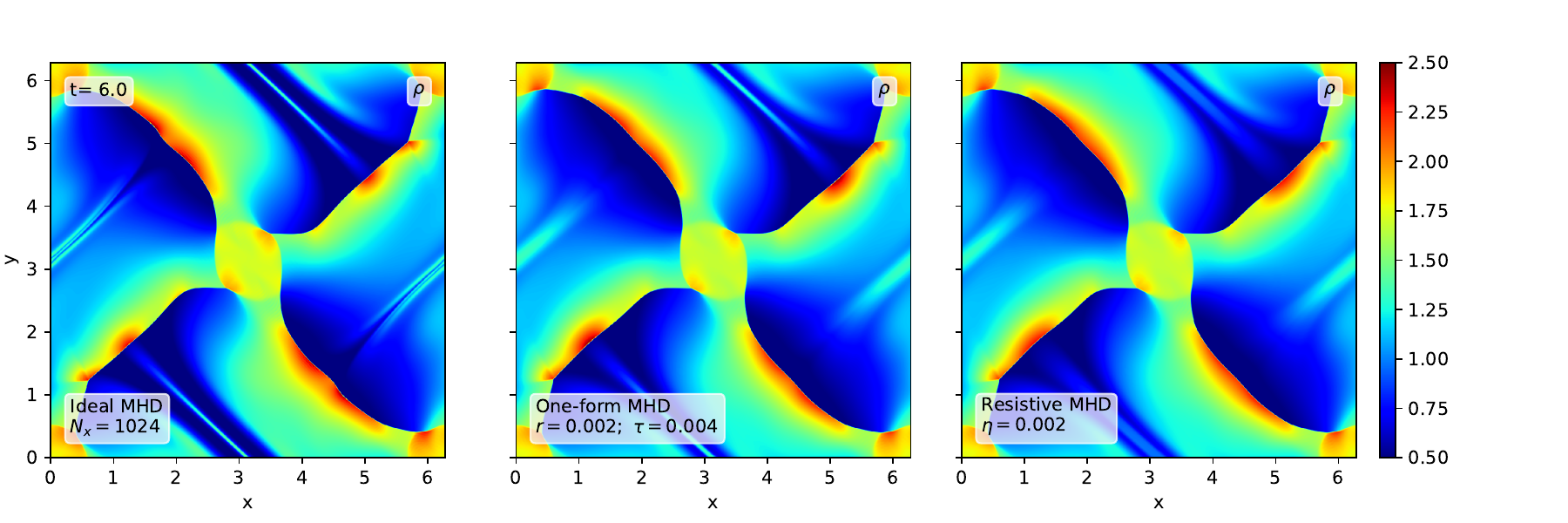}\\
    \includegraphics[trim={0 0.2cm 1cm 0.5cm},clip,width=17cm]{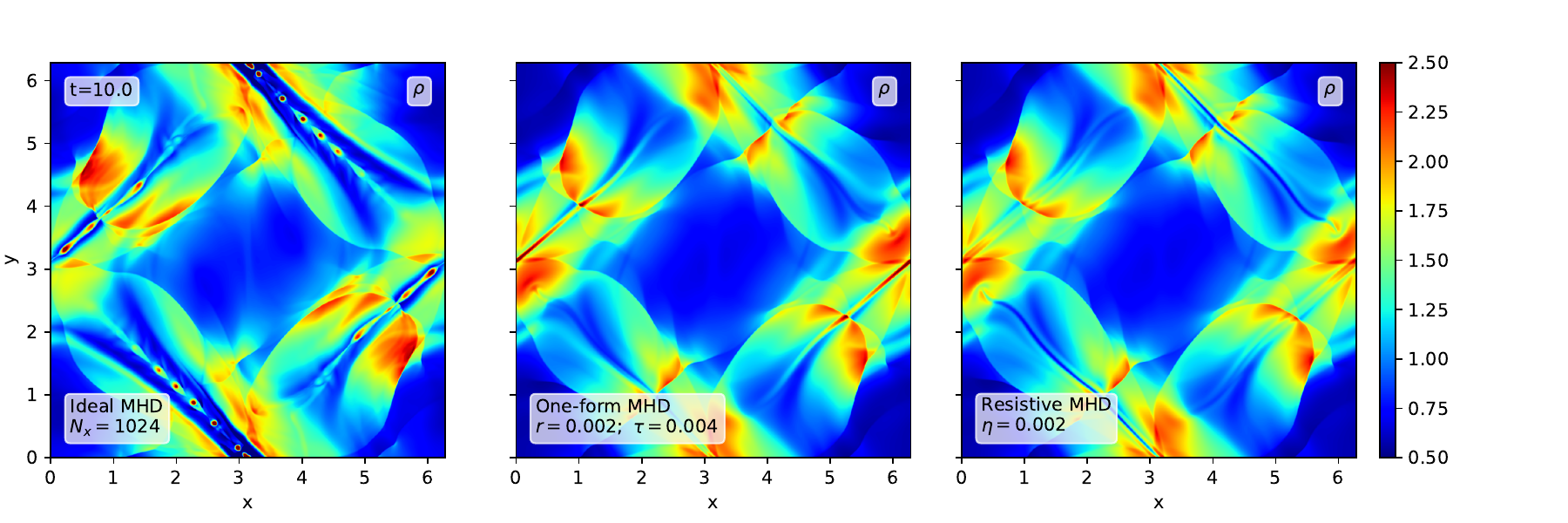}\\
    \caption{Density evolution in the relativistic Orszag-Tang vortex problem, comparing ideal, one-form and traditional resistive MHD for equivalent initial conditions.}
    \label{fig:OT}
\end{figure*}
\begin{figure}[t] 
    \centering
    \includegraphics[trim={0 0 0 0},clip,width=0.48\textwidth]{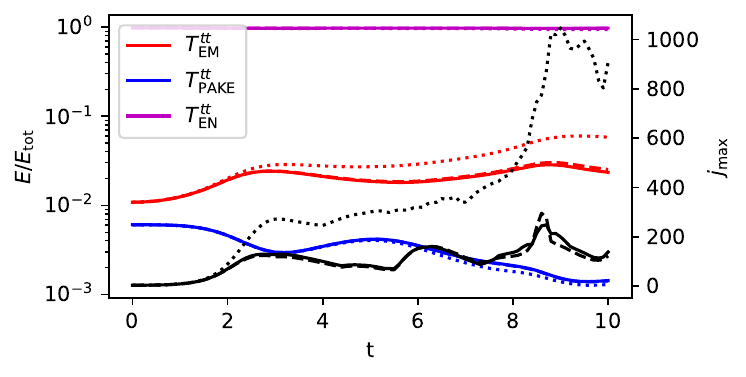}
    \caption{Energy evolution of the Orszag-Tang vortex problem. Solid lines: one-form MHD, dashed lines: traditional resistive MHD, dotted lines: ideal MHD (at resolution of $1024^2$ gridpoints). In black (right y-axis), we show the maximum value of the current density magnitude for all three cases.\label{fig:OTenergetics}}
\end{figure}
  
The Orszag-Tang vortex \citep{OrszagTang1979} is a standard two-dimensional test problem for MHD codes where small-scale structure and current sheets develop after an initial ideal evolution of MHD.
With this test, we can assess the ability of the scheme to recover the (near-)ideal evolution for small dissipative coefficients. Although one-form MHD corresponds to ideal MHD at leading order, evolving the system for small dissipative corrections is nontrivial since the matrix (\ref{eq:mij}) becomes singular in the limit $r\to0, \tau\to 0$.  
The initial vector components of the Orszag-Tang vortex are given by
\begin{equation}
    \begin{aligned}
    \big(u^x, u^y\big)  
    &= \Big({-}\Gamma v_{\rm max} \sin(y), \Gamma v_{\rm max} \sin(x)\Big)\\
    \big(J^{tx},J^{ty}\big) 
    &= \Big({-}\sin(y), \sin(2x)\Big)\, 
\end{aligned}
\end{equation}
together with density $\rho=1$ and pressure $p=10$. We choose $v_{\rm max}=0.8$, $r=0.002, \tau=2 r$.
The comoving magnetic field is initialized under the assumption of negligible gradient terms such that $b^i = (J^{ti} + b^t u^i)/\Gamma$ and $b^t=J^{ti} u_i$. The doubly periodic computational domain is given by $x,y\in[0,2\pi]$ and resolved by $1024^2$ cells.  This test is executed with a timestep corresponding to a courant parameter of ${\rm CFL}=0.2$ and $\partial_i J^{it}=0$ preserving FCT algorithm.  
    
We compare the evolution in one-form MHD with both ideal MHD as well as a traditional resistive MHD simulations \citep{RipperdaBacchiniEtAl2019} with matched parameters. For the traditional resistive MHD simulation, we hence set the resistivity to $\eta=0.002$ and apply the first-second IMEX timeintegrator due to \cite{BucciantiniDel-Zanna2013}. All other numerical parameters are chosen identical to the one-form case.  While resistivity in ideal MHD simulation is formally zero, the finite grid resolution introduces dissipative properties which are often paraphrased as ``numerical resistivity'' \citep[see][for a detailed investigation of this effect]{GrehanGhosalEtAl2025}.
Figure \ref{fig:OT} compares the density 
at intermediate ($t=6$) and late ($t=10$) times of evolution.

As expected, the evolution on large (ideal) scales is very similar between all formulations.  It is particularly noteworthy that the strong shocks visible at $t=6$ are also very well recovered by one-form MHD despite the use of discretized gradients in fluxes and source terms.  
Differences are most noticeable in the diagonal current sheet (going through the point (4.0,5.5) at $t=6$), which is very sharp in ideal MHD and appears most diffused in resistive MHD. At late times, the current-sheets in ideal MHD become tearing-unstable and fragment into multiple plasmoids that are ejected from the sheets \citep[see e.g.][for further discussion of the onset of plasmoid instability in the Orszag-Tang vortex]{2020ApJ...900..100R}.

To quantitatively compare the dynamics, we compute the electromagnetic $T^{tt}_{\rm EM}$, particle-kinetic $T^{tt}_{\rm PAKE}$ and enthalpy $T^{tt}_{\rm EN}$ energy contributions \citep[see][for definitions]{McKinneyTchekhovskoy2012}. The total energy $T^{tt}=T^{tt}_{\rm EM} + T^{tt}_{\rm PAKE} + T^{tt}_{\rm EN}$ is conserved to machine precision in all formulations. Figure \ref{fig:OTenergetics} shows these contributions as function of time. Initially, the vortical velocity field performs work on the magnetic field, leading to an increase of $T^{tt}_{\rm EM}$ at the expense of $T^{tt}_{\rm PAKE}$.  At $t\simeq3$, the current density reaches a maximum and the system ``bounces back'' as the accumulated magnetic pressure pushes against the flow.  This behavior is identical between all three formulations.  However, in ideal MHD, the maximum current density continues to increase and the magnetic energy content remains larger compared to the cases with explicit dissipation. By contrast, the quantifications (energy contributions and maximum current density) in one-form and traditional resistive MHD are nearly indistinguishable throughout the entire evolution.  
Further quantifications for the Orszag-Tang vortex are provided in App.~\ref{sec:OT-details}.

\aj{\section{Discussion and perspectives}}

In this paper we have presented a novel approach to dissipative relativistic magnetohydrodynamics, namely one where the plasma is treated as a fluid with conserved strings that are diffusive due to the presence of resistivity.
In practical terms, this approach enables one to avoid using \Ampere's law and the related stiffness issues. To demonstrate the veracity of the new formulation, we have developed a causal numerical implementation where we consider an additional evolution equation for the comoving magnetic field $b^i$. This is done using a scheme based on \cite{Pandya_2021} that inverts the second order corrections added to the constitutive equations, which are necessary to maintain causality.  Investigating the front velocity, we provide a proof of the causality of the proposed system of equations. \rl{Lastly, we test the numerical scheme using shocktubes, Orszag-Tang vortex problem, and cylindrical explosions}. We find that the system of equations can be evolved stably using standard numerical techniques and the evolution proceeds very similar to traditional MHD.

\rl{In \cite{Wright_2019}, an approach to numerically solving resistive MHD was followed that shows similarities with the one considered in this work. In particular, in this work the stiff Ampere's law is perturbatively expanded so that it instead turns into a diffusive correction proportional to resistivity, which is precisely how resistivity enters into the one-form MHD formalism. However, one major difference with the approach considered in this work is that \aj{we have added a second-order term to the dissipative two-form current that contains resistivity as a first order correction. This term renders the one-form MHD equations causal in a way that is similar to traditional resistive MHD.}} \op{In contrast to \cite{Wright_2019}, our scheme therefore does not suffer from loss of stability for high resolutions and only requires the standard hyperbolic timestep $\Delta t \propto \Delta x/c$.  }

\op{While the front-velocities of the proposed system have been shown to remain bounded and real (upon appropriate choice of the relaxation time), a full investigation of hyperbolicity remains for future work.  It is worth pointing out that the standard set of resistive GRMHD equations ``with  Amp\`ere's law'' is usually attributed as ``hyperbolic'' \citep[e.g.][]{Komissarov2007, MignoneMattiaEtAl2019}, however as \citep[][]{SchoepeHilditchEtAl2018, HilditchSchoepe2019}  point out, it is only weakly hyperbolic which indicates an ill-defined Cauchy problem.  }

In the past decade, one-form MHD was proposed as a more natural and fundamental realisation of MHD that manifests its global symmetry structure. In particular, the one-form formulation casts all MHD equations into flux-conservative form, which has been anticipated to be better suited for numerical evolution. However, despite many formal developments over the years, progress in the practical implementation of one-form MHD has been limited. This paper is the first step in this direction and opens up an entirely new promising avenue for future exploration. 

A natural extension of this work is to consider the full set of dissipative transport inherent to MHD \cite{Armas20201}. While such task may appear daunting when combined with causal completions such as BDNK \cite{Armas_2022}, we expect to avoid costly nonlinear inversions from conservative to primitive variables and to directly obtain evolution equations for primitive variables \citep{Pandya_2021}. The establishment of such a robust numerical scheme would be suitable for studying a wider class of extreme astrophysical phenomena, such as black hole accretion and neutron star mergers. The resulting practical implementation would form the basis for numerical schemes aimed at studying other forms of relativistic extreme matter in which higher form symmetries are present \cite{Armas:2023tyx}, such as superfluids in neutron star cores. We plan to address these in the future.


\acknowledgments

We would like to thank Filippo Camilloni, Pavel Kovtun, Raphael Hoult, Martin Pessah and Luciano Rezzolla for helpful discussions. The authors are partly supported by the Dutch Institute for Emergent Phenomena (DIEP) cluster at the University of Amsterdam and JA via the DIEP programme Foundations and Applications of Emergence (FAEME). The work of AJ was partly funded by the European Union’s Horizon 2020 research and innovation programme under the Marie Skłodowska-Curie grant agreement NonEqbSK No. 101027527, and partly by DIEP.

\clearpage
\appendix 

\onecolumngrid

{\center{\large\bfseries Appendix}\par}

\vspace{0.5cm}
The structure of this appendix is as follows. In Sec.~\ref{sec:Alfven1}, we provide a covariant derivation of Aflven's theorem suitable to one-form symmetry which helps us understand how reconnection requires dissipative corrections. In Sec.~\ref{app:causality}, we compute the front velocity corresponding to the linearized waves of our model to verify that causality is upheld. In Sec.~\ref{eq:traditionalMHD} and~\ref{sec:connecting_higher_form}, we compute the front velocity of traditional and one-form MHD respectively in the ultra-relativistic limit and show that $\tau$ can be chosen in relation to $r_{\perp,\parallel}$ such that there is a precise match between the front velocities. This verifies that this second order term serves to restore the inherently causal nature of traditional MHD in this dual one-form description of MHD. In Sec.~\ref{eq:primitivevariablerecovery}, we elaborate the part of our numerical scheme involving primitive variable recovery. 
In Sec.~\ref{sec:OT-details}, we elaborate on the Orszag-Tang evolution, focusing on the curl of the magnetic field, which can be viewed as the out-of-plane current density.
\section{Covariant proof of Alfven's theorem}
\label{sec:Alfven1}

\begin{figure}[t]
    \centering
    \includegraphics[width=7cm]{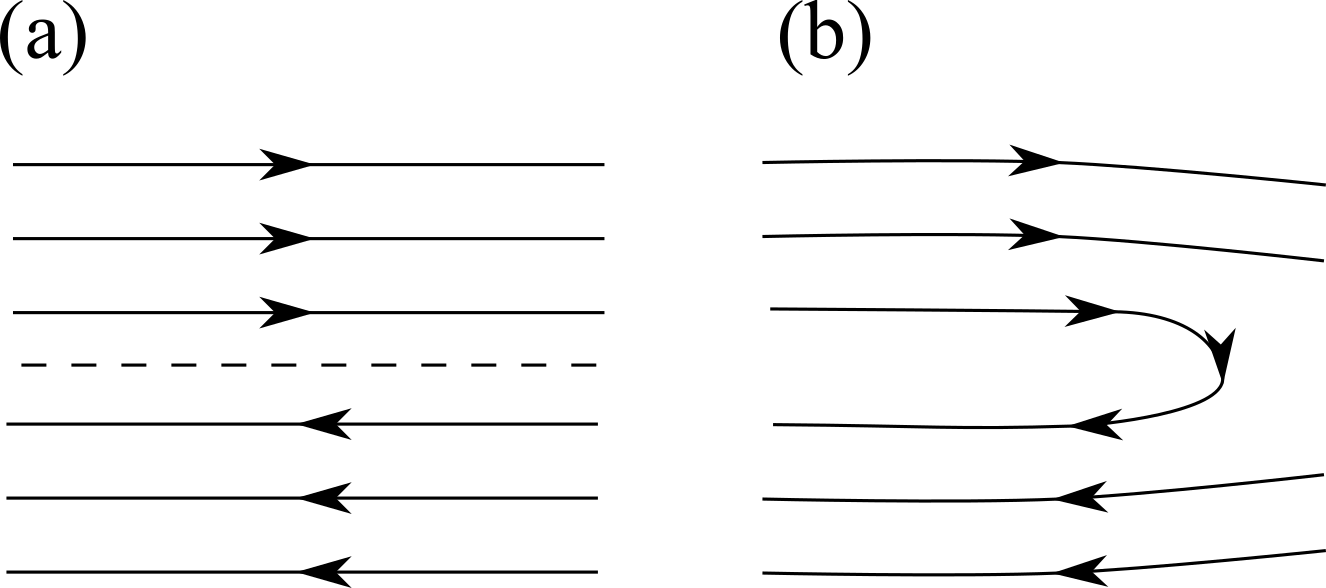} 
    \caption{A current sheet (dashed line) undergoing reconnection. The arrows are magnetic field lines.}
    \label{fig:single-column}
\end{figure}

In this section we explain how Alfven's theorem~\cite{Davidson_2001} can be understood in the one-form language for relativistically covariant surfaces. We start by defining the magnetic flux across a two-dimensional spatial surface $\Sigma$ given as
\begin{align}
    \Phi[\Sigma]
    \equiv 
    \int_\Sigma J^{\mu\nu}
    \df S^\Sigma_{\mu\nu}
    = \int_\Sigma \half \epsilon^{\mu\nu\rho\sigma}F_{\rho\sigma}
    \df S^\Sigma_{\mu\nu},
\end{align}
where $\df S^\Sigma_{\mu\nu}$ denotes the volume element on $\Sigma$. In the language of higher-form symmetries, this is precisely the conserved string charge passing through $\Sigma$~\cite{iqbal2024jenalecturesgeneralizedglobal}. For example if we use Cartesian coordinates and $\Sigma$ is just the $x$-$y$ plane, this simply becomes $\int\df x\df y\sqrt{-g}\,{\bf B}^{z}$. Consider transporting the surface $\Sigma$ along the fluid velocity $u^\mu$ to another surface $\Sigma'$. In the process, the boundary $\dow\Sigma$ gets transported to $\dow\Sigma'$. We will call the $(d-p)$-dimensional surface between $\dow\Sigma$ and $\dow\Sigma'$ traced during this process as $\Gamma$. Now consider the volume $\cM$ enclosed between $\Sigma$, $\Sigma'$, and $\Gamma$. Using Gauss's theorem and one-form conservation, we have that
\begin{align}
    0 &= \int_{\cM} \nabla_\mu J^{\mu\nu} \df S^\cM_{\nu} \nn\\
    &= \int_\Sigma J^{\mu\nu} \df S^\Sigma_{\mu\nu}
    - \int_{\Sigma'} J^{\mu\nu} \df S^{\Sigma'}_{\mu\nu}
    + \int_\Gamma J^{\mu\nu} \df S^\Gamma_{\mu\nu}.
\end{align}
We have a minus sign in the $\Sigma'$ integration because we choose $\df S^{\Sigma}_{\mu\nu}$ and $\df S^{\Sigma'}_{\mu\nu}$ to have the same orientation. Since $\Gamma$ is generated by a transport along $u^\mu$, it follows that $u^\mu$ is one of the tangent vectors of $\Gamma$ and $u^\mu \df S^\Gamma_{\mu\nu}=0$. Using this in the equation above, we find
\begin{align}
    \Phi[\Sigma]
    - \Phi[\Sigma']
    = - \int_\Gamma J^{\mu\nu}_{(1)} \df S^\Gamma_{\mu\nu}.
\end{align}
In other words, the magnetic flux through $\Sigma$ and $\Sigma'$ is only different in the presence of dissipative corrections. If we take $\Sigma$ and $\Sigma'$ to be infinitesimally apart, $\Gamma$ approaches $\dow\Sigma$ and this gives rise to
\begin{align}
    \delta_u\Phi[\Sigma]
    = \int_{\dow\Sigma} 3J^{[\mu\nu}_{(1)} u^{\rho]} \df S^{\dow\Sigma}_{\mu\nu\rho},
\end{align}
integrated on the boundary of $\Sigma$. The fact that the right hand side of \eqref{eq:reconnection} becomes nonzero in the presence of dissipative corrections means that the magnetohydrodynamic fluid can display reconnection (see Fig.~\ref{fig:single-column}).

We can also write down an analogous statement for the ``ideal'' magnetic flux
\begin{align}
    \Phi_0[\Sigma]
    \equiv 
    \int_\Sigma 2u^{[\mu}b^{\nu]}
    \df S^\Sigma_{\mu\nu}.
\end{align}
However, note that this object is \emph{not} the true magnetic flux and also depends on the choice of hydrodynamic frame. For our $x$-$y$ plane example, this instead evaluates to $\int\df x\df y\sqrt{-g}\,\Gamma^2({\bf B}-{\bf v}\times{\bf E}-{\bf v}({\bf B}\cdot{\bf v}))^z$. This reduces to $\int\df x\df y\sqrt{-g}\,{\bf B}^z$ in the ideal case when ${\bf E}={\bf B}\times{\bf v}$. Nonetheless, using Gauss's law we find
\begin{align}
    \int_{\cM} \nabla_\mu \Big(2u^{[\mu}b^{\nu]}\Big) \df S^\cM_{\nu}
    &= \int_\Sigma 2u^{[\mu}b^{\nu]} \df S^\Sigma_{\mu\nu}
    - \int_{\Sigma'} 2u^{[\mu}b^{\nu]} \df S^{\Sigma'}_{\mu\nu}
    + \int_\Gamma 2u^{[\mu}b^{\nu]} \df S^\Gamma_{\mu\nu}~.
\end{align}
Using the conservation equation and the fact that  $u^\mu \df S^\Gamma_{\mu\nu}=0$, we obtain
\begin{align}
    \Phi_0[\Sigma] - \Phi_0[\Sigma']
    &= - \int_{\cM} \nabla_\mu J^{\mu\nu}_{(1)}\, \df S^\cM_{\nu}.
\end{align}
Once again, we see that the difference is only nonzero in the presence of dissipative corrections. The associated infinitesimal version is given as
\begin{align}  \label{eq:reconnection}
    \delta_u\Phi_0[\Sigma]
    &= \int_{\Sigma} 2\nabla_\mu J^{\mu[\nu}_{(1)} u^{\rho]}\, 
    \df S^\Sigma_{\nu\rho}.
\end{align}

\section{Causality of one-form MHD}
\label{app:causality}

\label{sec:causality}
In this section, we verify the causality of one-form MHD equations with the BDNK constitutive relations \eqref{eq:diffusiveterm} for the two-form current and keeping the remaining constitutive equations to be ideal as in \eqref{eq:idealconstitutiveequations1}. We start with fluctuations of an equilibrium state with
\begin{align} \label{eq:fluctuations}
    T = T_0 + \delta T  ~~ , \qquad
    \mu = \mu_0 + \delta\mu  ~~ , \qquad
    u^{\mu }  = \delta^\mu_t + \delta u^{\mu}  ~~, \qquad
    b^{\mu} = b_0\delta^\mu_z + \delta b^{\mu} ~~ ,
\end{align}
such that $\delta u^t = 0$ and $\delta b^t  = -\delta u^z$, and consider plane-wave perturbations of the form $\sim \exp(i k_{\mu} x^{\mu} )$ with
 \begin{align}
     k^{\mu}  = \Big(\omega , \kappa \sin(\theta)  ,  0  ,  \kappa \cos(\theta) \Big) ~~ .
 \end{align}
The dispersion relations are obtained by setting the determinant of the linearized equations of motion $H(\omega,\kappa, \theta)$ to zero. Due to rotational and parity invariance, the determinant factorizes as
 \begin{align}
     H (\omega , \kappa , \theta ) 
     = H_{\text{Alfven}} (\omega , \kappa , \theta ) 
     \cdot
     H_{\text{magnetosonic}} (\omega , \kappa , \theta ) ~~, 
 \end{align}
corresponding to the Alfven and magnetosonic waves respectively. We impose causality by obtaining the front velocity $W (\theta ) $ for each of these waves, with 
\begin{align}
W  (\theta ) =  \lim_{ \kappa \rightarrow \infty }  \frac{ \omega}{\kappa } ~~ ,
\end{align}
and imposing~\cite{Krotscheck1978}
\begin{align} \label{eq:causalityconstraints123}
    \text{Re}\,W(\theta )   \leq 1 ~~ , \qquad 
    \text{Im}\,W(\theta )   =  0  ~~ .    
\end{align}
We assume for simplicity that $r\equiv r_{\perp} = r_{\parallel}$, so that $J^{\mu \nu }$ is given by
\begin{align}  \label{eq:fullmodel}
    J^{\mu \nu}_{(1) }   
    &=  J^{\mu \nu }_{(0)}
    -  2  r   P^{\mu \rho } P^{\nu \sigma }   T \partial_{[ \rho } \!\left(\frac{b_{\sigma ] } }{T} \right) -   2 \tau  u^{[\mu}  \nabla_{\rho} J_{(0)}^{\nu]  \rho }   ~~  , 
\end{align}
where $P^{\mu \rho }  = \eta^{\mu \nu } + u^{\mu } u^{\nu} $. To account for temperature variations, we use the ideal gas relation
\begin{align}
    T \propto \frac{p}{\rho}~~ , 
\end{align}
where the proportionality coefficient drops out of \eqref{eq:fullmodel}. Let us first consider the Alfven channel, where the front velocities are given by
\begin{align}
     W_{\text{Alfven}}^2 (\theta ) = \frac{b_0^2  \cos^2 ( \theta )}{   b_0^2+w_0 } 
  + \frac{r}{\tau} ~~, 
\end{align}
which can be constrained to satisfy $ W_{\text{Alfven}}^2  \leq 1 $ for sufficiently large $\tau $. For the magnetosonic channel, obtaining $W_{\text{Alfven}}(\theta)$ requires solving the sixth order polynomial
\begin{subequations} \label{eq:sixthorderpolynomial}
\begin{align}
 A  +   B\,W_{\text{magnetosonic}}^2 (\theta) +  C\,W_{\text{magnetosonic}}^4 (\theta)
       +D\,W_{\text{magnetosonic}}^6 (\theta) 
     =0  ~~,
\end{align}
where $A,B,C,D$ are given by
\begin{align}
    A &=   -  (\hat \gamma -1)^3  b_0^4 p_0 r \cos^2(\theta)\sin^2(\theta ) ~~ , \\ 
    \begin{split}
        B  &= (\hat \gamma -1) \left(\hat \gamma ^2 p_0^3 r+(\hat \gamma -1) \hat \gamma  p_0^2 \left(\rho_0 r+b_0^2 \left(\tau \cos ^2(\theta )+r \left(2 \sin ^2(\theta )+1\right)\right)\right)\right) \\  
        &\qquad +(\hat \gamma -1)^2 \sin ^2(\theta ) \left(b_0^2 p_0 r \left(b_0^2 \left((\hat \gamma -1) \cos ^2(\theta )+\hat \gamma \right)+3 (\hat \gamma -1) \rho_0\right)+(\hat \gamma -1) b_0^4 \rho_0 r\right)  ~~ ,  
    \end{split} \\ 
    \begin{split}
        C &=-(\hat \gamma -1) p_0 \tau \left(b_0^2 \left(\hat \gamma  p_0 \left((\hat \gamma -1) \cos ^2(\theta )+1\right)+(\hat \gamma -1) \rho_0\right)+\hat \gamma  p_0 (\hat \gamma  p_0+(\hat \gamma -1) \rho_0)\right) \\
        & \qquad -r (\hat \gamma  p_0+(\hat \gamma -1) \rho_0) \left((\hat \gamma -1) b_0^2+\hat \gamma  p_0+(\hat \gamma -1) \right) \left((\hat \gamma -1) b_0^2 \sin ^2(\theta )+p_0\right)  ~~ ,
    \end{split} \\ 
    D &= p_0 \tau \Big(\hat \gamma p_0+(\hat \gamma -1) \rho_0\Big)
    \Big((\hat \gamma -1) b_0^2+\hat \gamma p_0+(\hat \gamma -1) \rho_0\Big) ~~ .
\end{align}
\end{subequations}
One can verify numerically that $W_{\text{magnetosonic}}$ is always real for thermodynamically consistent parameters. That is, the equations are hyperbolic. Furthermore, $W_{\text{magnetosonic}}^2 \leq 1 $ can always be satisfied for a sufficiently large value of $\tau$. Therefore, we see that our one-form MHD model is causal. The analysis can analogously be repeated for $r_\|\neq r_\perp$. 

\section{Front velocity of traditional resistive MHD in the ultra-relativistic limit }\label{eq:traditionalMHD}

In this section, we review the spectrum of traditional resistive relativistic MHD in the absence of mass density, which is related to ultra-relativistic one-form MHD with second order corrections that we discuss later in App.~\ref{sec:connecting_higher_form}. In this case, we can replace temperature variations by simply $\df T = T/(\epsilon + p)\,\df p$. The component of the \Ampere's law \eqref{eq:ampere} along $u^\mu$ can be used to fix the charge density $q = \nabla_\mu e^\mu - F^{\mu\nu} \dow_\mu u_\nu =\cO(\dow)$. The remaining spatial components of the \Ampere's law \eqref{eq:ampere}, together with the energy-momentum and one-form conservation equations in \eqref{eq:conservationlaws}, determine the dynamics of $e^\mu$, $b^\mu$, $u^\mu$, and $T$.
The constitutive relations are given as
\begin{subequations}  
    \begin{align}  \label{eq:stress}
    T^{\mu \nu }   
    &=  \Big( \epsilon  + p  +  b^2 \Big)  u^{\mu} u^{\nu} 
    +   \left( p  +  \frac12 b^2 \right) g^{\mu \nu}    
    -  b^{\mu } b^{\nu}   
    - 2 u^{( \mu}\epsilon^{\nu)\rho\sigma\lambda}  
    e_{\rho} u_{\sigma} b_{\lambda}  + \mathcal{O} (\partial^2 ) ~~ ,     \\ 
    \label{eq:Fequation}
    F^{\mu \nu }   &  =2 u^{[\mu} e^{\nu]}  -  \epsilon^{\mu \nu \rho \sigma  } u_{\rho } b_{\sigma  }  ~~  ,  \\ 
  \label{eq:Jequation}     J^{\mu \nu }   &  = 2 u^{[\mu} b^{\nu]}  +  \epsilon^{\mu \nu \rho \sigma  } u_{\rho } e_{\sigma  }~~, 
\end{align}
\end{subequations}
together with the charge current
\begin{align}
    J^{\mu}  =  \sigma_{\parallel} \hat b^{\mu} \hat b^{\nu}   e_{\nu}  +  \sigma_{\perp} \mathbb{B}^{\mu  \nu}   e_{\nu}   ~~ .  
\end{align}
In the Alfven channel, we find the damped modes and a pair of sound modes \cite{Hernandez_2017}
\begin{align}
   \omega  =  -i\sigma_{\parallel  } + \mathcal{O} (\kappa^2 )~~ ,\qquad 
   \omega  =  -i \sigma_{\perp }\frac{w_0+b_0^2}{w_0} + \mathcal{O} (\kappa^2  )  ~~,\qquad 
  \omega  =  \pm \kappa  \frac{b_0 \cos (\theta )}{\sqrt{w_0 + b_0^2}} + \mathcal{O} (\kappa^2  )  ~~ , 
\end{align}
where $w_0 = \epsilon_0+p_0$. In the magnetosonic channel, on the other hand, we find a damped mode and a pair of sound modes
\begin{align}
      \omega  = -i \sigma_{\perp }\frac{w_0+b_0^2}{w_0} + \mathcal{O} (\kappa^2)~~, \qquad 
    \omega = \pm v_{\pm^{\prime }} (\theta )   \kappa  + \mathcal{O} (\kappa^2  )   ~~  , 
\end{align}
with the speed given by 
\begin{align}  \label{speedofsound}
    v^2_{\pm^{\prime }} (\theta )   = \frac{b_0^2 \left( \frac{\partial p}{\partial \epsilon} \cos ^2(\theta )+1\right) + \frac{\partial p}{\partial \epsilon} w_0  \pm' \sqrt{\left( \frac{\partial p}{\partial \epsilon} 
 ( b_0^2 \cos ^2(\theta )  + w_0 )  +b_0^2\right)^2-4  \frac{\partial p}{\partial \epsilon} b_0^2 \cos ^2(\theta ) \left(b_0^2+w_0\right)}}{2 \left(b_0^2+w_0\right)} ~~ . 
\end{align}
Note that for the ideal gas equation of state \eqref{eq:EoS}, we have $\dow p/\dow\epsilon = \hat\gamma-1$.
The front velocities are given by
\begin{align}\label{eq:front-velocities}
    W_{\text{Alfven}} (\theta ) = \pm 1~~, \qquad 
    W_{\text{magnetosonic}} (\theta ) 
    = \left\{ \pm 1 , \pm \sqrt{ \frac{\partial p}{\partial \epsilon}} \right\}~~ , 
\end{align}
which means the fastest front velocity is luminal. 

\aj{\section{Transforming traditional resistive MHD to one-form MHD in the ultra-relativistic limit}
\label{eq:transform}
In this Appendix we show that at first order in gradients, one can transform the equations of traditional resistive MHD to those of one-form MHD. \aj{For simplicity, we will work in the ultra-relativistic limit.} Let us \aj{begin with} the equations of motion \aj{traditional resistive MHD, i.e.}
\begin{subequations}  \label{eq:relativisticMHD12}
\begin{align} \label{eq:noenergy123}
  \partial_{\nu} T^{\mu \nu }  &=   0   ~~ ,  \\   \label{eq:maxwellequation}
    \partial_{\nu} F^{\mu \nu}  &=  J^{\mu}   ~~  . 
    \end{align}    
We also have the Bianchi identity
\begin{align}
       \partial_{\nu} J^{ \mu \nu}  =  0 ~~ ,   
\end{align}
\end{subequations}
\aj{where $J^{\mu\nu}=\half\epsilon^{\mu\nu\rho\sigma}F_{\rho\sigma}$.}
We plug \eqref{eq:Fequation} into \eqref{eq:maxwellequation}, which leads to
\begin{align}  \label{eq:substitutedequations}
    - \partial_{\nu} \Big(
    \epsilon^{\mu \nu \rho \sigma  } u_{\rho} b_{\sigma  }
    \Big)  &= 
    \sigma_{\parallel} \hat b^{\mu} \hat b^{\nu}   e_{\nu}  +  \sigma_{\perp} \mathbb{B}^{\mu  \nu}   e_{\nu}  
  + \mathcal{O} ( \partial^2 )     ~~  . 
\end{align}
Projecting \eqref{eq:substitutedequations} \aj{along $u^\mu$, $b^\mu$, and the transverse directions, we find}
\begin{subequations}
    \begin{align}
    \epsilon^{\mu \nu \rho} \partial_{\mu} u_{\nu} b_{\rho }   
    &= \mathcal{O} ( \partial^2 )      ~~  ,  \\
    - \frac{1}{  \sigma_{\perp }}      \mathbb{B}^{\mu}_{~\lambda} 
    \epsilon^{\lambda\nu\rho\sigma}  \partial_{\nu} \left(u_{\rho}  b_{\sigma} \right) 
    =     
    - \frac{1}{  \sigma_{\perp }} \left(     \hat E^{\mu\nu}  \hat b^{\rho}  X_{\nu\rho}         
    +  |b|   \hat E^{\mu\nu}   u^{\rho}    D_{\nu\rho}   \right)
    &= \mathbb{B}^{\mu\nu}e_\nu  
    +\mathcal{O} ( \partial^2 )  ~~  , \label{eq:electricfieldthing} \\ 
    -   \frac{1}{\sigma_{\parallel}}   \hat E^{\mu \nu } 
    \partial_{\mu}  b_{\nu }  
    &=  \hat b_{ \mu}   e^{\mu}     +   \mathcal{O} ( \partial^2 )   ~~   , 
\end{align}
\end{subequations}
where we introduced
\begin{align} 
     \hat E^{\mu\nu} = \epsilon^{\mu \nu \rho \sigma }  u_{\rho } \hat b_{\sigma}  ~~   , \qquad 
     X_{\mu\nu}   =     P^{\rho}_{\mu}  P^{\sigma}_{\nu}   \partial_{[\rho} 
   b_{\sigma] }    ~~ , \qquad 
           D_{\mu\nu}  =  \partial_{(\mu}   u_{\nu) }~~.
   \end{align}
Using energy-momentum balance, it is possible to relate the term $  \mathbb{B}^{\lambda\mu} u^{\nu} D_{\mu \nu }$ in \eqref{eq:electricfieldthing} to $  \mathbb{B}^{\lambda\mu} b^{\nu}  X_{\mu \nu }$. Specifically, considering energy-momentum balance projected transversely to the fluid velocity and magnetic field yields
\begin{align}
\begin{split}
 \mathbb{B}_{\lambda\nu}   \Big( ( \epsilon  + p    + b^2   )    u_{\mu}  D^{\mu \nu }      +  \partial^{\nu}  p  +     b_{\mu} X^{\nu \mu } \Big)   &  =  \mathcal{O} ( \partial^2 )  ~~ , 
  \end{split}
\end{align}
where we \aj{have} used that $ u^{\nu}   \partial_{\mu} u_{\nu}  =0 $. Finally, we obtain
\begin{align}  \label{eq:simpleidentity}
    \mathbb{B}^{\mu \rho } u^{\nu} D_{\rho \nu }  =    -    \frac{1 }{ \epsilon + p   + b^2 } \mathbb{B}^{\mu \rho }   (  b^{\sigma } X_{\rho \sigma }  +   \partial_{\rho}  p  )  +  \mathcal{O} ( \partial^2 )  ~~ , 
\end{align}
so that plugging \eqref{eq:simpleidentity} into~\eqref{eq:electricfieldthing} leads to
\begin{align}
\begin{split}
    e^{\mu}     
    &= - \frac{1}{2\sigma_\parallel} \hat b^\mu \hat E^{\rho\sigma} X_{\rho\sigma}     
    -   \frac{\epsilon  + p }{ \sigma_{\perp} ( \epsilon  + p +  b^2   )  }  
    \hat E^{\mu\nu}     \hat  b^{\rho}   X_{\nu\rho}     
    +   \frac{ |b|  }{ \sigma_{\perp} ( \epsilon  + p +  b^2   )  }  \hat E^{\mu\nu}         \partial_{\nu}  p  +  \mathcal{O}  ( \partial^2  )    ~~  .  
\end{split}
\end{align}
We can then substitute this into \eqref{eq:stress} and \eqref{eq:Jequation} to find
\begin{subequations}  \label{eq:TJcurrent}
    \begin{align}
    \begin{split}
            T^{\mu \nu }   
            &  =    ( \epsilon  + p  +  b^2   )  u^{\mu} u^{\nu} 
            +   ( p  +  b^2  /2    )    \eta^{\mu \nu}    
            - b^{\mu } b^{\nu}  
            -   \frac{ 
            2 ( \epsilon  + p  )  }{\sigma_{\perp}  ( \epsilon  + p +   b^2   )   }   
            u^{( \mu} \mathbb{B}^{\nu )\rho}    b^{\sigma} X_{\rho\sigma}      
            \\    
            &\qquad   
            +  \frac{2 b^2   }{\sigma_{\perp}  ( \epsilon  + p +   b^2   )   }   
            u^{( \mu} \mathbb{B}^{\nu )  \rho}  \partial_{\rho} p      + \mathcal{O}  ( \partial^2  )    
 ~~    , 
     \end{split}
\\ 
    \begin{split}
         J^{\mu \nu }   &  =   2 u^{[\mu } b^{ \nu ] }   - \frac{1}{\sigma_\parallel}  \mathbb{B}^{\mu \rho }  \mathbb{B}^{\nu \sigma  }     X_{\rho\sigma}   
         -      \frac{2 ( \epsilon + p )  }{\sigma_{\perp} (\epsilon + p + b^2 )} \mathbb{B}^{\rho[\mu}   \hat b^{\nu]}    \hat b^{\sigma}  X_{\rho\sigma}     \\ 
     &        +       \frac{ 2 |b|   }{\sigma_{\perp} (\epsilon + p + b^2 )} \mathbb{B}^{\rho [\mu}   \hat b^{\nu]}   \partial_{\rho}  p    
     +  \mathcal{O}  ( \partial^2  ) ~~    .          \end{split}
    \end{align}
\end{subequations}
We can write this as
    \begin{subequations}  \label{eq:TJcurrent123}
    \begin{align}  \label{eq:Tcontribution}
    \begin{split}
            T^{\mu \nu }   &  =    ( \epsilon  + p  +  b^2   )  u^{\mu} u^{\nu} +   ( p  +  b^2  /2    )    \eta^{\mu \nu}    - b^{\mu } b^{\nu}  -   \frac{ 
 2 ( \epsilon  + p  )  }{\sigma_{\perp}  ( \epsilon  + p +   b^2   )   }   u^{( \mu} \mathbb{B}^{\nu )  \rho}    b^{\sigma}       X'_{\rho\sigma}       + \mathcal{O}  ( \partial^2  )    
 ~~    , 
     \end{split}
\\ 
\begin{split}
         J^{\mu \nu }   &  =   2 u^{[\mu } b^{ \nu ] }   -  \left[ \frac{1}{\sigma_\parallel}  \mathbb{B}^{\mu\rho}  \mathbb{B}^{\nu \sigma}        +   \frac{2 ( \epsilon + p )  }{\sigma_{\perp} (\epsilon + p + b^2 )} \mathbb{B}^{\rho[\mu}   \hat b^{\nu]}    \hat b^{\sigma} \right]  X'_{\rho\sigma}     +  \mathcal{O}  ( \partial^2  ) ~~   .         \end{split}
    \end{align}
    \end{subequations}
    where
    \begin{align}
         X^{\prime }_{\rho\sigma}   = P^{\mu}_{\rho}  P^{\nu}_{\sigma}   T  \partial_{[\mu}  \left( 
   \frac{ b_{\nu] } }{ T}    \right) ~~ , 
    \end{align}
and $ T $ is the temperature which obeys 
\begin{align}
 \frac{1}{  T }   \partial_{\mu}   T   =  \frac{1}{\epsilon + p }  \partial_{\mu}  p  
\end{align}
Lastly, let us perform a frame transformation which removes the dissipative contribution in \eqref{eq:Tcontribution}. We take
\begin{align}
    u^{\mu} \rightarrow      u^{\mu} 
    +  \frac{\epsilon  + p    }{\sigma_{\perp}  ( \epsilon  + p +   b^2   )^2    }
    \mathbb{B}^{\mu  \nu}    b^{\rho}       X^{\prime }_{\nu\rho}     ~~ , 
\end{align}
so that \eqref{eq:TJcurrent123} turns into
 \begin{subequations}  \label{eq:TJcurrent123}
    \begin{align}  
    \begin{split}
            T^{\mu \nu }   &  =    ( \epsilon  + p  +  b^2   )  u^{\mu} u^{\nu} +   ( p  +  b^2  /2    )    \eta^{\mu \nu}    - b^{\mu } b^{\nu}     + \mathcal{O}  ( \partial^2  )    
 ~~    , 
     \end{split}
\\ 
\begin{split}
         J^{\mu \nu }   &  =   2 u^{[\mu } b^{ \nu ] }   -  \left[ \frac{1}{\sigma_\parallel}  \mathbb{B}^{\mu \rho }  \mathbb{B}^{\nu \sigma}        +   \frac{2 ( \epsilon + p )^2  }{\sigma_{\perp} (\epsilon + p + b^2 )^2 } \mathbb{B}^{\rho[\mu}   \hat b^{\nu]}    \hat b^{\sigma} \right]  X^{\prime }_{\rho\sigma}     +  \mathcal{O}  ( \partial^2  ) ~~   .         \end{split}
    \end{align}
    \end{subequations}
Comparing this to \eqref{eq:generalcurrent} we find the relations
\begin{align}
\label{eq:mapping}
 \left(   \frac{w   + b^2 }{w }  \right)^2    r_{\perp} = \frac{1}{\sigma_{\perp}}  ~~ , ~~ r_{\parallel}  = \frac{1}{\sigma_{\perp}  } ~~ . 
\end{align}}

\section{Front velocity of one-form MHD in the ultra-relativistic limit }\label{sec:connecting_higher_form}
In this section we show that in absence of mass density, it is possible to make a specific choice for $r_{\perp }$, $r_{\parallel}$ so that the front velocities of dual MHD with second order corrections coincides with that of traditional MHD. This means that, just like traditional resistive MHD, this version of dual MHD is causal under all circumstances. We take
\begin{align} \label{eq:relationcorrect}
   \frac{w   + b^2 }{w }   r_{\perp } =  r_{\parallel }   =  \tau ~~ . 
\end{align} 
We will again consider the fluctuations \eqref{eq:fluctuations}.
For the Alfven channel, we find a single damped mode and a pair of sound modes
\begin{align}
    \omega = -\frac{i}{\tau  }+ \mathcal{O} (\kappa^2  ), \qquad
    \omega = \pm \kappa  \frac{b_0\cos (\theta )}{\sqrt{w_0+b_0^2}} + \mathcal{O} (\kappa^2  )  ~~ , 
\end{align}
On the other hand, in the magnetosonic channel we find two damped modes and a pair of sound modes
\begin{align}
    \omega =  - \frac{i}{\tau } + \mathcal{O} (\kappa^2  ) ~~ , \qquad 
    \omega =  - \frac{i}{\tau } + \mathcal{O} (\kappa^2  ), \qquad 
    \omega = \pm v_{\pm^{\prime }} (\theta )   \kappa + \mathcal{O} (\kappa^2  )   ~~  , 
\end{align}
with the sound speed equal to \eqref{speedofsound}. Furthermore, the front velocities in this model coincide with those found in the traditional resistive formulation in \cref{eq:front-velocities}.

As for the dampings, let us consider the mapping at first order between traditional and dual MHD which lead to the relation of \eqref{eq:mapping}. Consistent with this mapping it follows that the conductivities of traditional resistive obey
\begin{align} \label{eq:relationcorrect12}
   \frac{w   + b^2 }{w }   \sigma_{\perp } =  \sigma_{\parallel }   =  \frac{1}{\tau}  ~~ . 
\end{align} 
We thus find that also the dampings coincide, although traditional MHD has two damped modes in the Alfven channel and one damped mode in the magnetosonic channel, whereas in the dual formulation there are two damped modes in the magnetosonic channel and one damped mode in the Alfven channel.

\section{Primitive variable recovery}
\label{eq:primitivevariablerecovery}

In this section, we give the details of primitive variable recovery for our numerical implementation. We use two complementary schemes that we outline below.

\subsection{\texttt{3D-Gamma-xi-bt} scheme}

The first scheme attempts to solve for three unknowns $x_A =\big(\Gamma,\xi\equiv\Gamma^2 (\epsilon + p ), b^t\big)$ using a system of scalar equations obtained from the constitutive relations for $\big(T^{ti}b_i, T^{ti} T^t_{~i}, T^{tt}\big)$. Hence we must find the simultaneous roots of the non-linear system
\begin{subequations} \label{eq:polynomials}
  \begin{align} 
  F_1(\Gamma,\xi,b^t) 
  &= b^t\Big(\xi + (\Gamma^2-1) {\bf b}^2 - \Gamma^2(b^t)^2 \Big) - T^{ti} b_i ~~ ,  \\
  F_2(\Gamma,\xi,b^t) 
  &= \lb 1 -\frac{1}{\Gamma^2}\rb 
    \Big(\xi + \Gamma^2 {\bf b}^2 - \Gamma^2(b^t)^2\Big)^2
    -  (b^t)^2 \bigg( (2\Gamma^2-1){\bf b}^2 + 2\xi - 2\Gamma^2(b^t)^2 \bigg) 
    -  T^{ti}T_{~i}^t~~,  \\
  F_3(\Gamma,\xi,b^t) 
  &= \xi 
  + \lb\Gamma^2-\half\rb {\bf b}^2 
  - \lb\Gamma^2+\half\rb (b^t)^2
  - p - T^{tt } ~~ , 
\end{align}
\end{subequations}
where ${\bf b}^2=b^ib_i$, which is done either with a three-dimensional Newton-Raphson or Newton-Krylov scheme. The pressure $p$ is determined by the equation of state as
\begin{align}
  p = \frac{\hat{\gamma}-1}{\hat{\gamma}} \frac{\xi-\Gamma \rho^t}{\Gamma^2} ~~ . 
\end{align}
Once $\Gamma$, $b^t$, and $p$ are known, $u^i$ and $\rho$ can be obtained using $T^{ti}$ and $\rho^t$ as
\begin{align}
    u^i
    = \frac{\Gamma}{\xi + \Gamma^2 {\bf b}^2 - \Gamma^2(b^t)^2}
    \Big( b^t b^i + T^{ti}\Big)~~, \qquad 
    \rho 
    = \frac1\Gamma \rho^t~~.
\end{align}

The Jacobian needed for the Newton-Raphson scheme is
\begin{subequations}
    \begin{align}
 J_{AB}  =     \frac{\partial}{\partial x_B } F_A  ~~ , 
\end{align}
and reads
\begin{align}
\begin{split}
& J_{AB}  = \\ 
& 
\begin{pmatrix}
    2\Gamma\,b^t \Big({\bf b}^2 - (b^t)^2\Big)
    & b^t 
    & (\Gamma^2-1){\bf b}^2 -3\Gamma^2 (b^t)^2 +\xi \\
    A 
    & 2 \Big( (\xi+\Gamma^2 {\bf b}^2) \left(1-\frac{1}{\Gamma^2}\right)-\Gamma^2 (b^t)^2 \Big) 
    & 2 b^t \Big((1-2\Gamma^4){\bf b}^2
    + 2\Gamma^2\left(\Gamma^2+1\right)(b^t)^2 - 2\Gamma^2\xi\Big) \\
    2 \Gamma \Big({\bf b}^2-(b^t)^2\Big) -  \partial_\Gamma p  & 1-\partial_\xi p & -b^t \left(2 \Gamma^2+1\right) 
     \end{pmatrix}, 
     \end{split}
\end{align}
where 
\begin{align}
 A  =  2 \Gamma \left((b^t)^4 -{\bf b}^4
 + 2 \Gamma^2 \Big({\bf b}^2-(b^t)^2\Big)^2 \right)
 + 4 \Gamma \xi \Big({\bf b}^2-(b^t)^2\Big)
 +\frac{2 \xi^2}{\Gamma^3} ~~ . 
\end{align}
\end{subequations}

\subsection{\texttt{3Dui} scheme}

For the second scheme, we use the expression for the relativistic momentum $T^{it}$ to formulate the primitive variable recovery as a root finding problem for the equations
\begin{align}\label{eq:3dui}
    F^i(u^i) 
    = \rho^t u^i
    + \Gamma  u^i \left(\frac{\hat{\gamma}}{\hat{\gamma}-1} p 
    + {\bf b}^2
    - \frac{1}{\Gamma^2} ({\bf b}\cdot{\bf u})^2
    \right) - \frac{{\bf b}\cdot{\bf u}}{\Gamma} b^i - T^{i t}~,
\end{align}
where $\Gamma(u^i) = \sqrt{1+u^i u_i}$ and the pressure follows from $T^{tt}$ according to
\begin{align}
    p = \frac{1}{\frac{\hat{\gamma}}{\hat{\gamma}-1} \Gamma^2 -1} 
    \left(T^{tt} - \Gamma \rho^t 
    - \lb \Gamma^2 - \half\rb {\bf b}^2 
    + \lb \Gamma^2 + \half\rb (b^t)^2\right)~~.
\end{align}
Equation (\ref{eq:3dui}) is solved using a Newton-Krylov algorithm which rapidly convergences after $\approx 3-4$ iterations in our numerical experiments.  In practice, this second ``\texttt{3Dui}'' scheme shows less inversion failures compared to the highly non-linear ``\texttt{3D-Gamma-xi-bt}'' scheme and is therefore preferred.  In the tests reported in this work, the ``\texttt{3Dui}'' inversion has operated without failure.

\section{Orszag-Tang evolution}
\label{sec:OT-details}
\begin{figure}[t] 
    \centering
    \includegraphics[trim={0 0 1cm 0.5cm},clip,width=16cm]{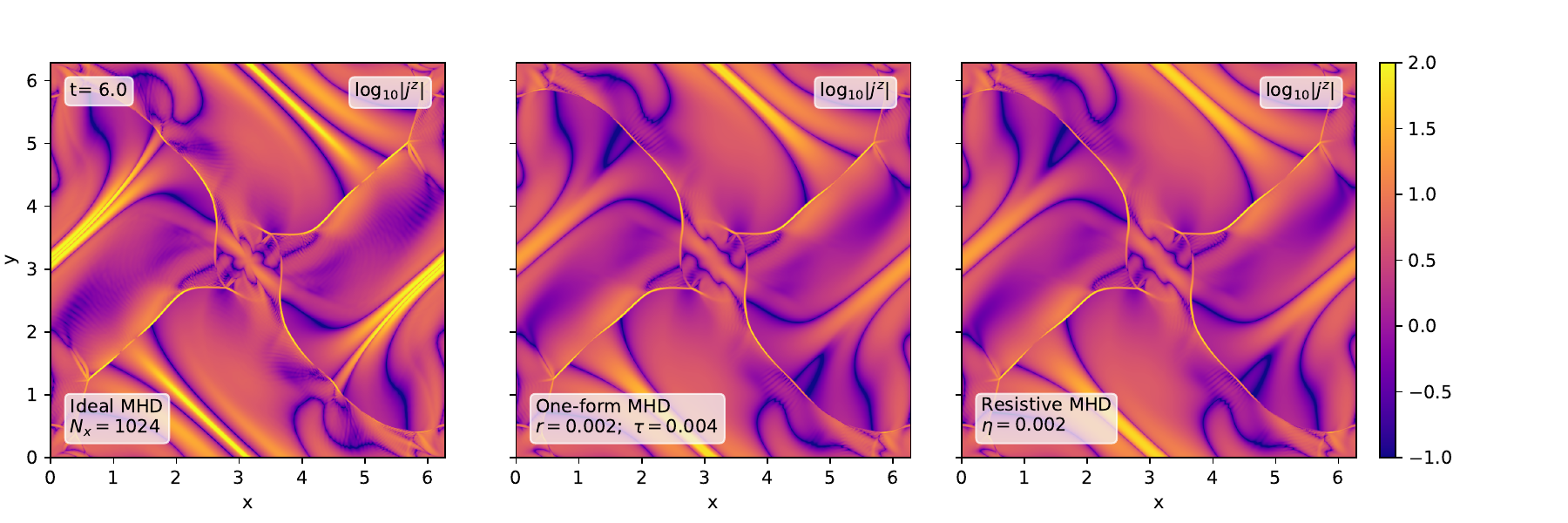} 
    \includegraphics[trim={0 0 1cm 0.5cm},clip,width=16cm]{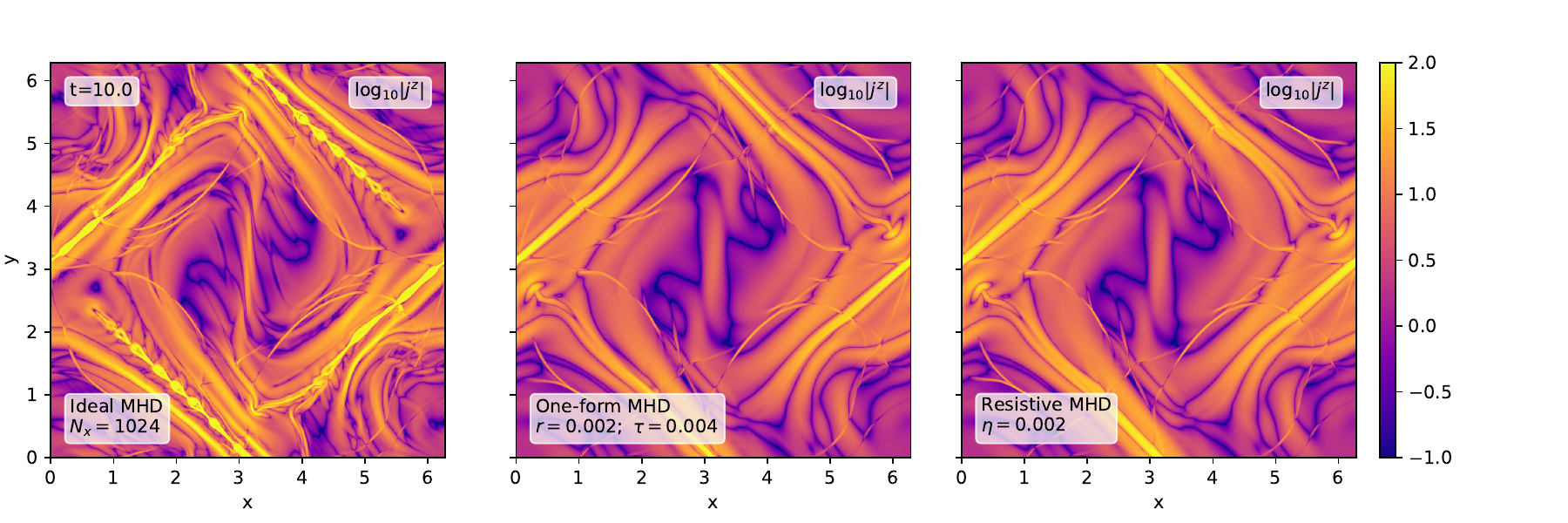} 
    \caption{Current evolution in the Orszag-Tang vortex problem, comparing ideal-, one-form and resistive relativistic MHD for the same initial conditions. }
    \label{fig:OT1}
\end{figure}

\begin{figure}[t] 
    \centering
    \includegraphics[trim={0 0 0 0},clip,width=0.4\textwidth]{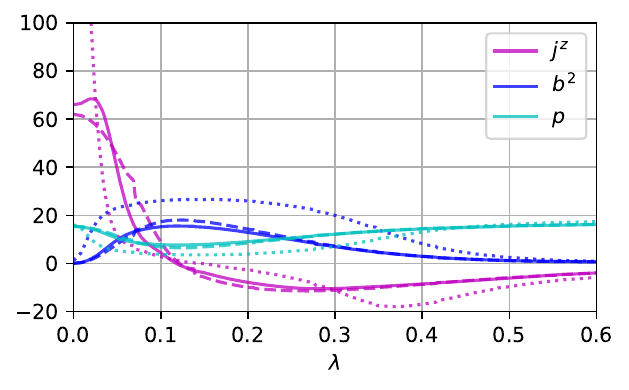}
    \caption{Cut across a current-sheet in the Orszag-Tang vortex problem for higher form MHD (solid lines), resistive MHD (dashed lines) and ideal MHD (dotted lines). We show the co-moving magnetic field magnitude $b^2$, gas-pressure $p$ and out-of-plane current density $j^z$ as function of the affine parameter along the cut $\lambda$.  The cases with dissipation show good qualitative agreement in the vicinity of the current sheet.  All formulations asymptote to a similar upstream ideal solution for large values of $\lambda$.  
    Snapshot taken at $t=6$, diagonal cut through points $A=(\pi,2\pi),\,B=(0,\pi)$.
    }
    \label{fig:OTcut}
\end{figure}

In this section we provide additional details on the Orszag-Tang evolution. Figure \ref{fig:OT1} compares a proxy for the out-of-plane current-density $j^z$ at intermediate ($t=6$) and late ($t=10$) times of evolution.  We define $\mathbf{j}=\nabla \times \mathbf{B}$ for the ideal and resistive runs and $\mathbf{j}=\nabla \times \mathbf{J}^{t}$ in higher form MHD.  
\newline 
A closer look at the dynamics in the vicinity of the current sheet is given in Figure \ref{fig:OTcut} where we cut across the upper current sheet (going through the point $(\pi,2\pi)$ in the upper panel of Figure \ref{fig:OT}).  The dissipative solutions show very similar properties regarding their balance of gas- and magnetic pressure as well as peak and full-width half-maximum of the current density.  All solutions approach each other in the ``ideal'' upstream region.  
\onecolumngrid\    

\end{document}